\documentclass[10pt,twocolumn,twoside]{IEEEtran}
\usepackage{amsmath,amsfonts}
\usepackage{amssymb}
\usepackage{multirow}
\usepackage{array}
\usepackage[caption=false,font=normalsize,labelfont=sf,textfont=sf]{subfig}
\usepackage{textcomp}
\usepackage{stfloats}
\usepackage{url}
\usepackage{verbatim}
\usepackage{graphicx}
\usepackage{cite}
\usepackage{cases} 
\usepackage[ruled]{algorithm2e}
\usepackage{setspace} 
\usepackage{amscd,dsfont,texdraw,tikz,verbatim,color,multicol} 
\usepackage{pdfsync}

\newtheorem{theorem}{\bf Theorem}[section]
\newtheorem{lemma}{\bf Lemma}[section]
\newtheorem{definition}{\bf Definition}[section]
\newtheorem{remark}{\bf Remark}[section]

\newtheorem{proposition}{\bf Proposition}[section]

\newtheorem{assume}{\bf Assumption}[section]

\newtheorem{problem}{\bf Problem}

\begin{document}

\title{Data-Driven Mean Field Equilibrium Computation in Large-Population LQG Games}

\author{Zhenhui Xu, Jiayu Chen, Bing-Chang Wang,~\IEEEmembership{Senior Member,~IEEE}, Tielong Shen,~\IEEEmembership{Member,~IEEE}
\thanks{Zhenhui Xu and Jiayu Chen are with School of Engineering, Institute of Science Tokyo, Tokyo, Japan (e-mail: xu.z.al@m.titech.ac.jp; jiayuc@mail.ustc.edu.cn).}
\thanks{Bing-Chang Wang is with School of Control Science and Engineering, Shandong University, Jinan, China (e-mail: bcwang@sdu.edu.cn).}
\thanks{Tielong Shen is with Faculty of Science and Engineering, Sophia University, Tokyo, Japan (e-mail: tetusin@sophia.ac.jp).}}

\maketitle

\begin{abstract}
This paper presents a novel data-driven approach for approximating the $\varepsilon$-Nash equilibrium in continuous-time linear quadratic Gaussian (LQG) games, where multiple agents interact with each other through their dynamics and infinite horizon discounted costs. The core of our method involves solving two algebraic Riccati equations (AREs) and an ordinary differential equation (ODE) using state and input samples collected from agents, eliminating the need for a priori knowledge of their dynamical models. The standard ARE is addressed through an integral reinforcement learning technique, while the nonsymmetric ARE and the ODE are resolved by identifying the drift coefficients of the agents' dynamics under general conditions. Moreover, by imposing specific conditions on models, we extend the learning-based approach to approximately solve the nonsymmetric ARE. Numerical examples are given to demonstrate the effectiveness of the proposed algorithms.
\end{abstract}

\begin{IEEEkeywords}
Nash equilibrium, linear quadratic Gaussian game, infinite horizon control, integral reinforcement learning
\end{IEEEkeywords}

\section{Introduction}\label{sec:introduction}
The mean field game (MFG) theory has gained significant attention as a tool to study Nash equilibrium in stochastic differential games involving a large number of noncooperative agents. Initially proposed in \cite{huang2007large,lasry2007mean}, the theory has been widely investigated \cite{bensoussan2013mean,carmona2018probabilistic,achdou2020introduction,luo2020multiagent,manjrekar2019mean,bakshi2019mean}. The core idea is to decouple each agent's individual decision from the behaviors of others by computing mass-effect quantities offline. Among various types of MFGs, linear quadratic mean field games (LQ-MFGs) have garnered particular attention due to their closed-form solution and theoretical tractability \cite{huang2006large,huang2007large,huang2010large,firoozi2022lqg,huang2021linear1,huang2019linear,huang2021linear,ma2020linear,zhang2021indefinite}. Following the Nash certainty equivalence (NCE) principle, it has been shown in \cite{huang2006large,huang2007large} that the solution to a finite player game, referred to as $\varepsilon$-Nash equilibrium, can be found by determining a best response to the optimal mean-field state that is consistent with the aggregate quantity. \cite{huang2010large} further challenged the major-minor case by the NCE method, and \cite{firoozi2022lqg,huang2021linear1} established the equivalency results of the $\varepsilon$-Nash equilibrium obtained via the NCE approach with those obtained via probabilistic approach and master equations, respectively. Meanwhile, \cite{huang2019linear} adopted the direct approach and investigated an asymptotic solvability problem, which is further extended to the controlled diffusion and indefinite weights setting and major-minor setting, in \cite{huang2021linear} and \cite{ma2020linear}, respectively. Moreover, we refer the reader to various studies of LQ-MFGs with different models and cost functions, such as risk-sensitive games\cite{moon2016linear}, games with common noise\cite{tchuendom2018uniqueness}, and games with delay systems \cite{huang2018linear}. 
 
Although MFG theory has demonstrated practical usefulness in various fields, analytically solving the coupled Hamilton-Jacobi-Bellman equation and Fokker-Planck-Kolmogorov equation, even when simplified to ARE and ODE in the LQ case, is still a challenging task. To overcome these difficulties, numerous model-based numerical methods have been developed and comprehensively discussed in the literature \cite{yin2010learning,cardaliaguet2018mean,hadikhanloo2019finite,fouque2020deep,cacace2021policy,carmona2021convergence,carmona2022convergence}. Meanwhile, the emergence of reinforcement learning techniques, coupled with the increasing affordability of sensing and computing resources, has driven growing interest in data-driven strategy design within the control systems community. In particular, integral reinforcement learning (IRL) has been proved effective for solving optimal control problems where system dynamics are partially or fully unknown. Instead of relying on an explicit dynamical model, IRL uses observed system trajectories and accumulated integrals to improve control policies. Originally introduced for LQ regulator problems in partially unknown linear systems \cite{vrabie2009neural}, IRL has since been extended to completely unknown linear systems \cite{jiang2012computational,lee2012integral}, nonlinear systems \cite{lee2014integral,xu2020model}, and stochastic systems \cite{li2022stochastic,xu2025mean}.  To develop mean-field reinforcement learning (MFRL) methods, how to deal with collective behavior in the environment with large number of agents lies at the heart of many studies. Early approaches to mean-field reinforcement learning, such as the first-order mean field approximation in \cite{yang2018mean} and the inverse reinforcement learning technique in \cite{yang2018deep}, primarily focusing on learning the mean-field game model without explicitly making decisions. More recent work, including the simulator-based Q-learning introduced in \cite{guo2019learning} and model-free algorithm developed in \cite{mishra2023model}, explore joint learning and decision-making processes. Additionally, methods such as fictitious play, which averages distributions of agents, and online mirror descent, which aggregates Q-functions, are thoroughly reviewed in \cite{cardaliaguet2017learning,perrin2020fictitious,perrin2021mean} and \cite{hadikhanloo2017learning,perolat2021scaling,anahtarci2023q}, respectively.

While many of these approaches are effective in settings with finite state and input sets, there are few results available for continuous state and input spaces, with \cite{fu2019actor,uz2020approximate,uz2020reinforcement,uz2023reinforcement,mao2022mean} being some notable exceptions. Specifically, \cite{fu2019actor} proposed an actor-critic algorithm for computing Nash equilibria in mean-field games with linear-quadratic dynamics, which was shown to converge to a Nash equilibrium under certain conditions. \cite{uz2020approximate} extended this work to non-stationary mean-field games using fixed-point iterations and actor-critic algorithms, while \cite{uz2020reinforcement}  studied Q-learning in this context. In addition, \cite{uz2023reinforcement} extended previous results on non-stationary mean-field games to the case of multiple populations, and \cite{mao2022mean} developed a function approximation approach based on neural networks for efficiently computing resource allocations close to a Nash equilibrium. Although there have been significant advances in understanding the discrete-time case of MFRL, there still exists an open issue in the development of numerical methods for continuous-time MFGs, which provides the motivation for the present study.

This paper proposes a novel data-driven methodology for determining the $\varepsilon$-Nash equilibrium in large-population LQG games. In this setting, the $\varepsilon$-Nash equilibrium is composed of both feedback and feedforward components. The feedback component requires solving an ARE to derive the feedback gain matrix, while the feedforward component involves solving another ARE to obtain the feedforward gain matrix and an ODE to determine the mean field stat trajectory. The interconnected nature of agents presents significant challenges in computing the $\varepsilon$-Nash equilibrium without \textit{a priori} knowledge of dynamical models. Specifically, the coupling terms in the dynamics and cost functions introduce two main technical difficulties. First, the coupling term in the dynamics obstructs the extension of the IRL technique to solve the feedback ARE. Second, these terms make the feedforward ARE asymmetric, increasing the complexity of solving this equation in the presence of unknown dynamics. To address these challenges, we integrate the system transformation, IRL techniques, and the identification methods to develop two data-driven algorithms. The key innovations of our approach are as follows:
 
1) A system transformation-based IRL method is developed to obtain the feedback matrix gain of the mean field equilibrium. A crucial step is applying a system transformation to overcome obstacles that hinder the extension from model-based to model-free iteration. Utilizing a unified dataset, an identification-based method is further developed to solve the nonsymmetric ARE and compute the mean field state trajectory, thus obtaining the feedforward part of the equilibrium.
 
2) For scenarios with specific conditions in the coupling terms of the dynamics and cost functions, the
nonsymmetric ARE is shown to be equivalent to solving an
indefinite ARE. This equivalence facilitates the development of a data-driven iterative equation to replace the identification-based method. It is worth noting that the off-policy iterations exhibit quadratic convergence, and the system parameters are computed only once under relaxed conditions compared to the persistence excitation condition required for the iteration processes. 

3) A new framework for solving general LQ-MFG problems is proposed. It consists of two phases: data collection and data-driven computation. This framework avoids the coupling between the strategy update and mean field approximation during the iteration, which is common in most existing algorithms. 

Our algorithms can solve more general LQG game problems compared to the previous method (\!\!\cite{xu2023mean}). The problem considered in \cite{xu2023mean} is a special case of the one we address in this paper. Moreover, unlike \cite{xu2023mean}, our approach avoids direct involvement of stochastic variables in the computation of integrals during off-policy iterations, thereby reducing computational complexity.

The rest of the paper is organized as follows.  Sec. \ref{sec2} presents the problem description.  Sec. \ref{sec3} introduces our data-driven method for computing the $\varepsilon$-Nash equilibrium. In Sec. \ref{sec4}, we focus on a special case and develop a new IRL method for determining the $\varepsilon$-Nash equilibrium. Sec. \ref{sec5} provides simulation examples to demonstrate the effectiveness of the proposed approach. Finally, conclusions are drawn in Sec. \ref{sec6}.
 
{\bf{Notation}:}
The following notations will be used. For a family of $\mathbb{R}^n$-valued random variables $\{x(\tau),\tau\!\geq0\}$, $\sigma(x(\tau),\tau\!\leq\! t)$ is the $\sigma$-algebra generated by these random variables; $\mathbb{S}^n$ denotes the set of $n\times n$ symmetric matrices; For a matrix $R\in\mathbb{S}^n$, $R>0$ denotes a positive definite matrix; For $v\in\mathbb{R}^m$ and $P\in\mathbb{R}^{m\times m}$, $\|v\|_P^2=v^{\mathrm{T}}Pv$; $\mathrm{col}(A)$ denotes the $m n$-dimensional vector formed by stacking the columns of $A\in\mathbb{R}^{m\times n}$ on top of one another; For $P \in \mathbb{S}^{n}$ and $x\in\mathbb{R}^n$, $\mathrm{colm}({P}) = [p_{11},2p_{12},\cdots,2p_{1n},p_{22},2p_{23},\cdots,2p_{n-1,n},p_{nn}]^{\mathrm{T}}$$\in$$\mathbb{R}^{\frac{1}{2}n(n+1)}$ and $\mathrm{colv}(x)=[x_1^2,x_1x_2,\cdots,x_1x_n, x_2^2,\cdots,x_{n-1}x_n, x_n^2]^{\mathrm{T}}\in\mathbb{R}^{\frac{1}{2}n(n+1)}$.

\section{Problem formulation}\label{sec2}
We consider a population of $N$ agents denoted as $\mathcal{A}=\{\mathcal{A}_i,1\leq i\leq N\}$, where $\mathcal{A}_i$ represents the $i$-th agent. Each agent's state process, represented by $x_i(t)$, is governed by the following stochastic differential equation (SDE):
\begin{equation}\label{sys1}
\mathrm{d}x_i(t)  = (Ax_i(t)+Gx_{(N)}(t) +Bu_i(t))\mathrm{d}t+D\mathrm{d}w_i(t), 
\end{equation}
where $x_{(N)}=\frac{1}{N}\sum_{k=1}^Nx_k$ denotes the average state of the population. The state $x_i$ and input $u_i$ are $n$-dimensional and $m$-dimensional vectors, respectively. The initial states $x_i(0)$, $1\leq i\leq N$, are mutually independent and identically distributed with $\mathbb{E}[x_i(0)]=\bar{x}_0$, and have finite second moment. The noise processes $\{w_i,1\leq i\leq N\}$ are $N$ independent standard $d$-dimensional Brownian motions and are also independent of the initial states. The constant matrices $A,B,G$, and $D$ all have compatible dimensions. 

Let $u_{-i}\!=\!(u_1,\!\cdots\!,u_{i-1},u_{i+1},\!\cdots\!,u_N\!)$ be the control inputs of all agents except $\mathcal{A}_i$. The cost function of $\mathcal{A}_i$ is given by
\begin{align}
\!\!\!J_i\big( u_i,u_{-i} \big) \!= \!\mathbb{E} \Big[ \int_0^{\infty}\!\!\! e^{-\rho\tau} \big(\|x_i- \Gamma x_{(N)}\|_{Q}^{2} + \|u_i\|_{R}^2 \big) \mathrm{d}\tau \Big],\label{J}
\end{align}
 where $\rho>0$ is a discount factor, $Q\geq0$ and $R>0$ have compatible dimensions, and $\Gamma\in\mathbb{R}^{n\times n}$. 

 The admissible decentralized control policy set of $\mathcal{A}_i$ is $\mathcal{U}_{i}=\big\{u_i:u_i(t) \text{ is adapted to } \mathcal{F}_t^i, \mathbb{E}\!\left[\int_0^{\infty}e^{\!-\rho \tau}\|u_i(\tau)\|^{2}\mathrm{d}\tau\right]<\infty\big\}$, where $\mathcal{F}_t^i=\sigma(x_i(\tau),\tau\leq t)$, $t\geq0$, $i=1,\cdots,N$. 

 \begin{definition}[\cite{huang2007large} ]
 A set of control policies $u_j^o\in\mathcal{U}_j$, $1\leq j\leq N$, for $N$ agents is called an $\varepsilon$-Nash equilibrium with respect to the costs $J_j$, $1\leq j\leq N$, if there exists $\varepsilon\geq0$ such that for any fixed $1\leq i\leq N$, we have 
 \begin{equation}
 J_i(u_i^o,u_{-i}^o)\leq J_i(u_i,u_{-i}^o)+\varepsilon,
 \end{equation}
 when any alternative $u_i\in\mathcal{U}_i$ is applied by $\mathcal{A}_i$.
 \end{definition}

To establish the $\varepsilon$-Nash equilibrium, the resolution involves solving the following AREs
\begin{numcases}{}
\rho P_{11}= P_{11}A+A^{\mathrm{T}}P_{11}-P_{11}BR^{-1}B^{\mathrm{T}}P_{11}+Q \label{HJB1}\\
\rho P_{12}= P_{12}(A+G)+A^{ \mathrm{T}}P_{12}-P_{12}BR^{-1}B^{\mathrm{T}}P_{12}\notag\\
~~~~~~~~~+Q-Q\Gamma,\label{HJB2}
\end{numcases} 
such that  the matrices $A-\rho/2I_n-BR^{-1}B^{\mathrm{T}}P_{11}^{*}$ and $A+G-\rho/2 I_n-BR^{-1}B^{\mathrm{T}}P_{12}^{*}$ are Hurwitz. To facilitate the analysis, define the $2n\times 2n$ block matrices as
\begin{equation*}
\begin{aligned}
&\mathcal{H}_1\triangleq\left[\begin{array}{cc}A-\frac{\rho}{2}I_n & -BR^{-1}B^{\mathrm{T}}\\-Q  &-A^{\mathrm{T}}+\frac{\rho}{2}I_n\end{array}\right],\\
&\mathcal{H}_2\triangleq\left[\begin{array}{cc}A+G-\frac{\rho}{2}I_n & -BR^{-1}B^{\mathrm{T}}\\Q\Gamma-Q &-A^{\mathrm{T}}+\frac{\rho}{2}I_n\end{array}\right].
\end{aligned}
\end{equation*}
The following conditions are needed:
\begin{enumerate}
\item[(A1)] The pair $(A-\rho/2I_n,B)$ is stabilizable and the pair $(A-\rho/2I_n,Q^{1/2})$ is observable.
\item[(A2)] The eigenvalues of $\mathcal{H}_2$ are strong $(n,n)$ c-splitting and the associated $n$-dimensional stable invariant subspace is a graph subspace.
\end{enumerate}
For further details regarding the concepts of strong c-splitting and graph subspaces, refer to \cite[page 10]{huang2019linear}.

\begin{proposition}[\cite{kleinman1968iterative}]
Let Assumption (A1) be satisfied, then equation (\ref{HJB1}) has a unique solution $P_{11}^*>0$ such that $A-\rho/2I_n-BR^{-1}B^{\mathrm{T}}P_{11}^{*}$ is Hurwitz.
\end{proposition}
\begin{proposition}[\cite{huang2019linear}]
Let Assumption (A2) be satisfied, then equation (\ref{HJB2}) has a unique solution $P_{12}^*\in\mathbb{R}^{n\times n}$ such that $A+G-\rho/2 I_n-BR^{-1}B^{\mathrm{T}}P_{12}^{*}$ is Hurwitz.
\end{proposition}
\begin{remark}
Note that the maximal stabilizing solution of (\ref{HJB1}) (or (\ref{HJB2})) can be put into one-to-one correspondence with a certain invariant subspace of the block matrix $\mathcal{H}_1$ (or $\mathcal{H}_2$). Furthermore, taking advantage of the Hermitian property the block matrix $\mathcal{H}_1$, the conditions for $\mathcal{H}_1$ (similar to those for $\mathcal{H}_2$ in (A2)) can be simplified to the stability and observability criteria of $(A-\rho/2I_n,B,Q^{1/2})$ as stated in (A1). 
\end{remark}

For simplicity in notation, let $K_1^*=R^{-1}B^{\mathrm{T}}P_{11}^*$ and $K_2^*=R^{-1}B^{\mathrm{T}}P_{12}^*$. Then, by using the fixed point approach \cite{huang2010large} (or the direct approach \cite{huang2019linear}), it gives that the set of control policies
\begin{equation}\label{uo}
u_i^o(t) = -K_1^{*}x_i(t)-(K_2^{*}-K_1^*)\bar{x}^*(t),~~1\leq i\leq N,
\end{equation}
constitute the $\varepsilon$-Nash equilibrium, where the corresponding mean field state $\bar{x}^*(t)$ satisfies the following ODE 
\begin{equation}\label{xbar}
\mathrm{d}{\bar{x}^*} = (A+G-BK_2^*)\bar{x}^*\mathrm{d}t,~~\bar{x}^*(0)=\bar{x}_0.
\end{equation}
   
However, solving the AREs (\ref{HJB1})-(\ref{HJB2}) and the mean field state system (\ref{xbar}) necessitates a priori knowledge of the coefficients $(A, B, G)$, thereby imposing constraints in numerous practical applications. Consequently, in the present paper, we dive into the challenge below.
\begin{problem}\label{prob1}
Develop a data-driven method to compute the $\varepsilon$-Nash equilibrium (\ref{uo}) of the LQG game (\ref{sys1})-(\ref{J}).
\end{problem}

\section{Data-driven $\varepsilon$-Nash equilibrium computation}\label{sec3}
In this section, we elaborate a solution to Problem \ref{prob1}. In particular, we construct this solution in three steps:  perform system transformations and data collection; utilize an IRL technique to solve for the feedback gain matrix; and  employ an identification technique, using the same dataset from the previous step, to determine the feedforward gain matrix and the mean field state trajectory.
 
\subsection{System transformation and data collection}
To proceed, define error variables and average variables 
\begin{align}
&\tilde{x}=\mathbb{E}[x_i-x_j],~~\tilde{u}=\mathbb{E}[u_i-u_j],~i\neq j,\label{xbar1def}\\
&\bar{x} = \mathbb{E}\left[\frac{1}{N}\sum\nolimits_{j=1}\nolimits^{N}x_{j}\right],~~\bar{u} = \mathbb{E}\left[\frac{1}{N}\sum\nolimits_{j=1}\nolimits^{N}u_j\right].\label{xbar2def}
\end{align}
By equation (\ref{sys1}), the dynamics of $\tilde{x}$ and $\bar{x}$ are given by 
\begin{align}
&\mathrm{d}\tilde{x}(t)=(A\tilde{x}(t)+B\tilde{u}(t))\mathrm{d}t,~~\tilde{x}(0) = {\bf0},\label{xbar1}\\
&\mathrm{d}\bar{x}(t) = ((A+G)\bar{x}(t)+B\bar{u}(t))\mathrm{d}t,~~\bar{x}(0) = \bar{x}_0.\label{xbar2}
\end{align}

Letting $T>0$ be the integration duration. Using the measurements of systems (\ref{xbar1}) and (\ref{xbar2}), we compute the following differences and integrals:
\begin{equation*}
\begin{aligned}
&\delta_{\tilde{x}\tilde{x}}^t \triangleq e^{-\rho(t+T)}\tilde{x}(t+T)\otimes\tilde{x}(t+T)-e^{-\rho t}\tilde{x}(t)\otimes\tilde{x}(t),\\
&\delta_{{\mathrm{colv}(\tilde{x})}}^t \triangleq e^{ -\rho(t+T)}{\mathrm{colv}(\tilde{x}(t + T))} - e^{\rho t} {\mathrm{colv}(\tilde{x}(t))},\\
&I_{\tilde{x}\tilde{x}}^t\!\triangleq\!\int_t^{t+T}\!\!\!\!\!\!e^{-\rho\tau}\tilde{x}(\tau)\!\otimes\!\tilde{x}(\tau)\mathrm{d}\tau,I_{\tilde{x}\tilde{u}}^t\!\triangleq\!\int_t^{t+T}\!\!\!\!\!\!e^{-\rho\tau}\tilde{x}(\tau)\otimes\tilde{u}(\tau)\mathrm{d}\tau, 
\end{aligned}
\end{equation*} 
\begin{equation*}
\begin{aligned}
&\delta_{\bar{x}\bar{x}}^t \triangleq e^{-\rho(t+T)}\bar{x}(t+T)\otimes\bar{x}(t+T)-e^{-\rho t}\bar{x}(t)\otimes\bar{x}(t),\\
&I_{\bar{x}\bar{x}}^t\!\triangleq\!\int_t^{t+T}\!\!\!\!\!\!e^{-\rho\tau}\bar{x}(\tau)\!\otimes\!\bar{x}(\tau)\mathrm{d}\tau,I_{\bar{x}\bar{u}}^t\!\triangleq\!\int_t^{t+T}\!\!\!e^{-\rho\tau}\bar{x}(\tau)\otimes\bar{u}(\tau)\mathrm{d}\tau.
\end{aligned}
\end{equation*}

Putting together $l$ samples for each calculation yields the following data-based matrices
\begin{equation}\label{matr1}
\left\{\begin{aligned}
&\Delta_{\tilde{x}\tilde{x}} = [\delta_{\tilde{x}\tilde{x}}^{t_1}, \cdots,\delta_{\tilde{x}\tilde{x}}^{t_l} ]^{\mathrm{T}}\in\mathbb{R}^{l\times n^2},\\
&\Delta_{{\mathrm{colv}(\tilde{x})}}= [\delta_{{\mathrm{colv}(\tilde{x})}}^{t_1}, \cdots,\delta_{{\mathrm{colv}(\tilde{x})}}^{t_l} ]^{\mathrm{T}}\in\mathbb{R}^{l\times\frac{1}{2}n(n+1)},\\
&\mathcal{I}_{\tilde{x}\tilde{x}} = [I_{\tilde{x}\tilde{x}}^{t_1}, \cdots,I_{\tilde{x}\tilde{x}}^{t_l} ]^{\mathrm{T}}\in\mathbb{R}^{l\times n^2}, \\
&\mathcal{I}_{\tilde{x}\tilde{u}} = [I_{\tilde{x}\tilde{u}}^{t_1}, \cdots,I_{\tilde{x}\tilde{u}}^{t_l} ]^{\mathrm{T}}\in\mathbb{R}^{l\times nm}, 
\end{aligned}\right.
\end{equation}
\begin{equation}\label{matr2}
\left\{\begin{aligned}
&\Delta_{\bar{x}\bar{x}} = [\delta_{\bar{x}\bar{x}}^{t_1}, \cdots,\delta_{\bar{x}\bar{x}}^{t_l} ]^{\mathrm{T}}\in\mathbb{R}^{l\times n^2},\\
&\mathcal{I}_{\bar{x}\bar{x}} = [I_{\bar{x}\bar{x}}^{t_1}, \cdots,I_{\bar{x}\bar{x}}^{t_l} ]^{\mathrm{T}}\in\mathbb{R}^{l\times n^2}, \\
&\mathcal{I}_{\bar{x}\bar{u}} = [I_{\bar{x}\bar{u}}^{t_1}, \cdots,I_{\bar{x}\bar{u}}^{t_l} ]^{\mathrm{T}}\in\mathbb{R}^{l\times nm},~~~~~~~~~~~~~~~~~~~
\end{aligned}\right.
\end{equation}
where $t_k = t_1+(k-1)T_s$ is the time point, $T_s$ is the sampling time, and $t_1$ is the initial time of data collection phase. For those matrices, we introduce assumptions:
\begin{assume}\label{A1}
There exists $l_1>0$ such that for all $l\geq l_1$,
\begin{equation}\label{rank1}
\mathrm{rank}\left(\left[\mathcal{I}_{\tilde{x}\tilde{x}},\mathcal{I}_{\tilde{x}\tilde{u}}\right]\right)=\frac{1}{2}n(n+1)+mn.
\end{equation}
\end{assume}
\begin{assume}\label{A2}
There exists $l_2>0$ such that for all $l\geq l_2$, 
\begin{equation}\label{rank2}
\mathrm{rank} \left(\mathcal{I}_{\bar{x}\bar{x}}\right)=\frac{1}{2}n(n+1).
\end{equation}
\end{assume}
\begin{remark}
\label{PE_condition}
Assumptions \ref{A1} and \ref{A2} resemble the persistent excitation (PE) condition (\cite{aastrom2008adaptive,jiang2012computational}), which are essential for ensuring the convergence of subsequent data-driven methods. To meet this condition, it is effective to introduce exploration noise into the input channel. Commonly used options for exploration noise in RL algorithm implementations include exponentially decreasing noise \cite{vamvoudakis2011multi}, random noise \cite{ahuja2016wellposedness}, and the sum of sinusoidal signals \cite{jiang2012computational}. 
\end{remark}
\begin{remark} 
In practice, the states and inputs are measured at discrete time instants with the sampling period $T_s$. Consequently, the continuous‐time integrals in the preceding derivations must be numerically approximated. A simple approach is the rectangle (Riemann) rule, although trapezoidal or higher‐order methods can also be used. For instance, when using the rectangle rule for the integral $I_{\tilde{x}\tilde{x}}^{t_l}$, one approximates it as $I_{\tilde{x}\tilde{x}}^{t_l}\approx T_s\sum_{k=l}^{l+N_t-1}e^{-\rho t_k}\tilde{x}(t_k)\otimes \tilde{x}(t_k)$, where $N_t = \frac{T}{T_s}$. Notice that $T_s$ is dictated by hardware constraints ({\sl e.g., sensor and processor capabilities}), while $T$ is a design parameter for the integration interval. Thus, to capture system behavior more accurately, one typically samples as fast as the hardware permits, ensuring a finer approximation of the integral.
 \end{remark}
\subsection{Data-based off-policy algorithm}
Note that the policy iteration, developed in \cite{kleinman1968iterative}, is an efficient algorithm to numerically approximate the solution of (\ref{HJB1}), and is stated in the following.
\begin{lemma}[\!\cite{kleinman1968iterative},\cite{pang2021robust}]\label{lem3.1}
Suppose there exists $K_1^0\in\mathbb{R}^{m\times n}$ such that $A-\rho/2I_n-BK_1^0$ is Hurwitz. Let $P_{11}^k\in\mathbb{S}^n$ be a solution of the following equation
\begin{equation}\label{PE}
\begin{aligned}
\rho P_{11}^k = & P_{11}^k(A-BK_1^{k-1})+(A-BK_1^{k-1})^{\mathrm{T}}P_{11}^k\\
&+\left(K_1^{k-1}\right)^{\mathrm{T}}RK_1^{k-1}+Q,
\end{aligned}
\end{equation}
and $K_1^{k}$ be recursively improved by 
\begin{equation}\label{PI}
K_1^k = R^{-1}B^{\mathrm{T}}P_{11}^k,~~k=1,2,\cdots.
\end{equation}
Then, the following properties hold:
\begin{itemize}
\item[1.] $\lim_{k\rightarrow\infty}P_{11}^k=P_{11}^*$ and $\lim_{k\rightarrow\infty}K_1^k=K_1^*$;
\item[2.] For any $\sigma\in(0,1)$, there exist $\delta_0(\sigma)>0$ and $c_0(\delta_0)>0$, such that for any $P_{11}^{k}\in\mathcal{B}_{\delta_0}(P_{11}^*)$, $\|P_{11}^{k+1}-P_{11}^*\|_F\leq c_0\|P_{11}^k-P_{11}^*\|_F^2$, where $\mathcal{B}_{r}(Z) \triangleq \{X\in\mathbb{R}^{n\times n}|\|X-Z\|_F<r\}$ and $\|\cdot\|_F$ is the Frobenius norm.
\end{itemize}
\end{lemma}

Based on the above result, we use the data-driven matrices (\ref{matr1}) to determine the $k$-th iterative solution of $P_{11}^*$ and $K_1^*$ by
\begin{equation}\label{pk}
 \left[\begin{array}{c}
{\mathrm{colm}({P}_{11}^k)}\\
\mathrm{col}(K_1^k)
\end{array}\right]=\left( \left( \Psi_1^k \right)^{\mathrm{T}}\Psi_1^k \right)^{ -1} ( \Psi_1^k )^{ \mathrm{T}}\Xi_1^k, 
\end{equation}
where
\begin{equation*}
\left\{\begin{aligned}
&\Psi_1^k = [\Delta_{{\mathrm{colv}(\tilde{x})}},-2\mathcal{I}_{\tilde{x}\tilde{x}}(I_n \otimes (K_1^{k-1})^{ \mathrm{T}}R) - 2\mathcal{I}_{\tilde{x}\tilde{u}}(I_n\!\otimes\! R)],\\
&\Xi_1^k=-\mathcal{I}_{\tilde{x}\tilde{x}}\mathrm{col}\left((K_1^{k-1})^{\mathrm{T}}RK_1^{k-1}+Q\right).
\end{aligned}\right.
\end{equation*}
The convergence of the iterative equation (\ref{pk}) is established in the following theorem.
\begin{theorem}\label{thm1}
Suppose there exists $K_1^0\in\mathbb{R}^{m\times n}$ such that $A-\rho/2I_n-BK_1^0$ is Hurwitz, and Assumption \ref{A1} holds. Then, the sequence $\{P_{11}^k,K_1^k\}_{k=1}^{\infty}$ generated by iteratively solving equation (\ref{pk}) possesses the following properties:

\begin{enumerate}
\item[1).] $\lim_{k\rightarrow \infty} P_{11}^k = P_{11}^* $ and $\lim_{k\rightarrow \infty} K_{1}^k = K_{1}^* $;
\item[2).]  For any $\sigma\in(0,1)$, there exist $\delta_0(\sigma)>0$ and $c_0(\delta_0)>0$, such that for any $P_{11}^{k}\in\mathcal{B}_{\delta_0}(P_{11}^*)$, $\|P_{11}^{k+1}-P_{11}^*\|_F\leq c_0\|P_{11}^k-P_{11}^*\|_F^2$;
\item[3).] $B=\left(P_{11}^{k}\right)^{-1}\!\!\left(K_{1}^{k}\right)^{\mathrm{T}}\!\!R$.
\end{enumerate}
\end{theorem} 
\begin{IEEEproof}
First, we demonstrate that under the assumption that $\left( \Psi_1^k \right)^{\mathrm{T}}\Psi_1^k$ is invertible, the solutions of (\ref{PE}) and (\ref{PI}) satisfy equation (\ref{pk}). Using the solution $P_{11}^k$ of equation (\ref{PE}), define a quadratic function 
\begin{equation*}
V_1(t,\tilde{x}(t))=e^{-\rho t}\tilde{x}(t)^{\mathrm{T}}P_{11}^k\tilde{x}(t).
\end{equation*}
Substitute the ODE (\ref{xbar1}) and the iterative equations (\ref{PE})-(\ref{PI}) into the time derivative of $V_1$ to obtain
\begin{equation}\label{dV1}
\begin{aligned}
\mathrm{d}V_1=&e^{-\rho t}\Big(2\left(\tilde{u}+K_{1}^{k-1}\tilde{x}\right)^{\mathrm{T}}RK_1^k\tilde{x}-\tilde{x}^{\mathrm{T}}\big(Q\\
&+\left(K_1^{k-1}\right)^{\mathrm{T}}RK_1^{k-1}\big)\tilde{x}\Big)\mathrm{d}t.
\end{aligned}
\end{equation}
Now, integrating both sides of this equation along the solution of (\ref{xbar1}) over the time interval $[t,t+T)$, it gives
\begin{equation}\label{iV1}
\begin{aligned}
&e^{-\rho(t+T)}\tilde{x}^{\mathrm{T}}(t+T)P_{11}^k\tilde{x}(t+T)-\!e^{-\rho t}\tilde{x}^{\mathrm{T}}(t)P_{11}^k\tilde{x}(t) \\
=&2\int_{t}^{t+T}e^{–\rho\tau}\left(\tilde{u}(\tau)+K_1^{k-1}\tilde{x}^{\mathrm{T}}(\tau)\right)RK_{1}^{k}\tilde{x}(\tau)\mathrm{d}\tau\\
&-\int_t^{t+T}\!\!\!\!\!e^{-\rho\tau}\tilde{x}^{\mathrm{T}}(\tau)\left(\left({K_1^{k-1}}\right)^{\mathrm{T}}RK_1^{k-1}+Q\right)\tilde{x}(\tau)\mathrm{d}\tau.
\end{aligned}
\end{equation}
By the Kronecker product representation and the data-based matrices (\ref{matr1}), the linear matrix form of equation (\ref{iV1}) can be rewritten as
\begin{equation}
\Psi_1^k\left[\begin{array}{c}
{\mathrm{colm}({P}_{11}^k)}\\
\mathrm{col}(K_1^k)
\end{array}\right]=\Xi_1^k,
\end{equation}
which implies that the solutions of equations (\ref{PE}) and (\ref{PI}) also satisfy equation (\ref{xbar1}).

Next, we remove the previous assumption by proving that $\Psi^k_1$ has full column rank for all $k\in\mathbb{N}_+$.
 
Assume $\Psi_1^kX={\bf0}$ for a nonzero $X=[X_1^{\mathrm{T}},X_2^{\mathrm{T}}]^{\mathrm{T}}$, where $X_1\in\mathbb{R}^{\frac{1}{2}n(n+1)}$ and $X_2\in\mathbb{R}^{nm}$. A symmetric matrix $Y\in\mathbb{S}^n$ can be uniquely determined by $\mathrm{colm}(Y)=X_1$, and a matrix $Z\in\mathbb{R}^{m\times n}$ can be determined by $\mathrm{col}(Z)=X_2$.

By equations (\ref{dV1}) and (\ref{iV1}), we obtain
\begin{equation}\label{p1eq1}
\Psi_1^k X = \mathcal{I}_{\tilde{x}\tilde{x}}\mathrm{col}(M) + 2\mathcal{I}_{\tilde{x}\tilde{u}}\mathrm{col}(N),
\end{equation} 
where $M=-\rho Y+Y(A-BK_1^{k-1})+(A-BK_1^{k-1})^{\mathrm{T}}Y +(K_{1}^{k-1})^{\mathrm{T}}N+N^{\mathrm{T}}K_1^{k-1}$ and $N=B^{\mathrm{T}}Y-RZ$.
Since $M\in\mathbb{S}^n$, we define $\mathcal{I}_{{\mathrm{colv}(\tilde{x})}}\!=\!\left[I_{{{\mathrm{colv}(\tilde{x})}}}^{t_1},\cdots,I_{{{\mathrm{colv}(\tilde{x})}}}^{t_l}\right]$ and obtain
\begin{equation}\label{p1eq2}
\begin{aligned}
\mathcal{I}_{\tilde{x}\tilde{x}}\mathrm{col}(M)=\mathcal{I}_{{\mathrm{colv}(\tilde{x})}}{\mathrm{colm}(M)},
\end{aligned}
\end{equation} 
where ${I_{\mathrm{colv}(\tilde{x})}^{t}}=\int_t^{t+T}{\mathrm{colv}(\tilde{x}(\tau))}\mathrm{d}\tau$.

Under Assumption \ref{A1}, the matrix $[\mathcal{I}_{{\mathrm{colv}(\tilde{x})}},2\mathcal{I}_{\tilde{x}\tilde{u}}]$ has full column rank. From $\Psi_1^k X={\bf0}$ and equations (\ref{p1eq1})-(\ref{p1eq2}), we get ${\mathrm{colm}(M)}={\bf 0}$ and $\mathrm{col}(N)={\bf 0}$. 

Thus, we have $-\rho Y+Y(A-BK_1^{k-1})+(A-BK_1^{k-1})^{\mathrm{T}}Y={\bf0}$, which implies $Y={\bf0}$ due to the Hurwitz property of the matrix $A-\rho/2 I-BK_1^{k-1}$. It follows that $Z={\bf0}$. Consequently, $X={\bf0}$, which contradicts our assumption. Therefore, $\Psi_1^k$ has full column rank for all $i\in\mathbb{N}_+$, ensuring the uniqueness of the solution to (\ref{pk}).

Hence, it can be immediately concluded that solving (\ref{pk}) is equivalent to solving (\ref{PE})-(\ref{PI}). Lemma \ref{lem3.1} directly guarantees the convergence of the sequence $\{P_{11}^{k},K_1^{k}\}_{k=1}^{\infty}$ and establishes its quadratic convergence rate, as stated in properties 1) and 2). Furthermore, property 3) follows from equation (\ref{PI}), thus completing the proof.  
\end{IEEEproof}

\subsection{Data-based identification}
To determine each row of $A$ and $A+G$, define $E_i\in\mathbb{R}^{n\times n}$ as the diagonal matrix with a value of one at the $i$-th entry on the diagonal and zeros elsewhere, and define $e_i\in\mathbb{R}^{n}$ as the vector with a value of one at the $i$-th entry and zero elsewhere. It is straightforward to see that $E_i=e_ie_i^{\mathrm{T}}$. Now, let $\kappa_1^i\in\mathbb{R}^n$ be the vector storing the $i$-th row of the matrix $A$, which can be computed using the subsequent result.
\begin{theorem}
Suppose the conditions of Theorem \ref{thm1} hold, then the constant matrix $A$ can be uniquely determined by
\begin{equation}\label{Ai}
\begin{aligned}
\kappa_1^i = \left(\left(\Psi_2^i\right)^{\mathrm{T}}\Psi_2^i\right)^{-1}\left(\Psi_2^i\right)^{\mathrm{T}}\Xi_2^i,~~i=1,2,\cdots,n,
\end{aligned}
\end{equation}
where $\Psi_2^i \!=\! 2\mathcal{I}_{\tilde{x}\tilde{x}}\left(e_i\otimes I_n\right)\in\mathbb{R}^{l\times n}$ and
$\Xi_2^i \!= \!\left(\Delta_{\tilde{x}\tilde{x}}+\rho \mathcal{I}_{\tilde{x}\tilde{x}} \right)\mathrm{col}\left(E_i\right)-2\mathcal{I}_{\tilde{x}\tilde{u}}\mathrm{col}\!\left(\!RK_1^k\!\left(P_{11}^k \right)^{\!-1}\! E_i\!\right)\! \in\!\mathbb{R}^{l}$.
\end{theorem}
\begin{IEEEproof}
 Under Assumption \ref{A1}, it is easy to get
 \begin{equation}
 \mathrm{rank}\left(\Psi_2^i\right)=n,~~i=1,2,\cdots,n,
 \end{equation}
 which implies the existence of the inverse of the matrix $\left(\Psi_2^i\right)^{\mathrm{T}}\Psi_2^i$, therefore, yielding the uniqueness result of the solution to equation (\ref{Ai}).
 Then, we complete the proof by showing that the constant matrix $A$ satisfies equation (\ref{Ai}). Define a new quadratic function
 \begin{equation}
 V_2(t,\tilde{x}(t)) = e^{-\rho t}\tilde{x}(t)^{\mathrm{T}}E_i\tilde{x}(t).
 \end{equation} 
 The time derivative of $V_2$ along (\ref{xbar1}) is 
 \begin{equation*}
 \mathrm{d}V_2 = e^{-\rho t}\left(-\rho\tilde{x}^{\mathrm{T}}E_i\tilde{x}+2\tilde{x}^{\mathrm{T}}\kappa_1^ie_i^{\mathrm{T}}\tilde{x}+2\tilde{u}^{\mathrm{T}}B^{\mathrm{T}}E_i\tilde{x}\right)\mathrm{d}t.
\end{equation*}
Taking the integration of the above equation over $[t,t+T)$ and changing the terms, it gives
\begin{equation*}
\begin{aligned}
&2\!\!\int_t^{t+T}\!\!\!\!\!\!e^{-\rho\tau}\tilde{x}^{\mathrm{T}}\kappa_1^ie_i^{\mathrm{T}}\tilde{x}\mathrm{d}\tau\!=\!e^{-\rho(t+T)}\tilde{x}(t\!+\!T)^{\mathrm{T}}E_i\tilde{x}(t\!+\!T)\!-\!e^{-\rho t}\!\\
&\times\tilde{x}(t)^{\mathrm{T}}\!E_i\tilde{x}(t)\!+\!\rho\! \!\int_{t}^{t+T}\!\!\!\!\!\!\!\!e^{-\rho\tau}\tilde{x}^{\mathrm{T}}\!E_i\tilde{x}\mathrm{d}\tau\!-\!2\!\!\!\int_{t}^{t+T}\!\!\!\!\!\!\!e^{-\rho\tau}\tilde{u}^{ \mathrm{T}}\!B^{ \mathrm{T}}\!E_i\tilde{x}\mathrm{d}\tau,
\end{aligned}
\end{equation*} 
which is equivalent to
\begin{equation*}
\!2\!\left(I_{\tilde{x}\tilde{x}}^t\right)^{\!\mathrm{T}}\!\!(e_i\otimes I_n)\kappa_1^i \!= \!\left(\delta_{\tilde{x}\tilde{x}}^t+\rho I_{\tilde{x}\tilde{x}}^t\right)^{\!\mathrm{T}}\!\!\!\mathrm{col}(E_i)-2\!\left(I_{\tilde{x}\tilde{u}}^t\right)\!\mathrm{col}\!\left( B^{\mathrm{T}}E_i\right).
\end{equation*}
Combining data from the $l$ sets and substituting $B^{\mathrm{T}}=RK_{1}^{k}(P_{11}^k)^{-1}$ into the above equation results in
\begin{equation}
\begin{aligned}
\Psi_2^i\kappa_1^i = \Xi_2^i,~~i=1,2,\cdots,n,
\end{aligned}
\end{equation}
which implies that the drift coefficient matrix $A$ satisfies equation (\ref{Ai}).
Hence, the proof is complete.
\end{IEEEproof}

Let $\kappa_2^i\in\mathbb{R}^n$ be the vector stored the $i$-th row of $A+G$.
\begin{theorem}
Suppose the conditions of Theorem \ref{thm1} and Assumption \ref{A2} hold, then the constant matrix $A+G$ can be uniquely determined by
\begin{equation}\label{Gi}
\begin{aligned}
\kappa_2^i=\left(\left(\Psi_3^i\right)^{\mathrm{T}}\Psi_3^i\right)^{-1}(\Psi_3^i)^{\mathrm{T}}\Xi_3^i,~~i=1,2,\cdots,n,
\end{aligned}
\end{equation}
where $\Psi_3^i \!=\! 2\mathcal{I}_{\bar{x}\bar{x}}\left(e_i\otimes I_n\right)\in\mathbb{R}^{l\times n}$ and
$\Xi_3^i \!= \!\left(\Delta_{\bar{x}\bar{x}}\!+\!\rho \mathcal{I}_{\bar{x}\bar{x}} \right)\mathrm{col}\left(E_i\right)-2\mathcal{I}_{\bar{x}\bar{u}}\mathrm{col}\!\left(\!RK_1^k\!\left(P_{11}^k \right)^{\!-1}\! E_i\!\right)\! \in\!\mathbb{R}^{l}$.
\end{theorem}
 
\begin{IEEEproof}
The full column rank condition of the matrix $\Psi_3^i$ under Assumption \ref{A2} guarantee the uniqueness of the solution to (\ref{Gi}). Subsequently, it is established that this unique solution is $\kappa_2^i$ by defining the function $V_3(t,\bar{x}(t))=e^{-\rho t}\bar{x}(t)^{\mathrm{T}}E_i\bar{x}(t)$, along with similar operators utilized in the above proof.
\end{IEEEproof}

Until now, we have determined the estimated values of the control parameters $(P_{11}^*,K_1^*)$ and the system parameters $(B,A,A+G)$. Using the certainty equivalence principle \cite{duncan1999adaptive}, we can solve $P_{12}^*$ from equation (\ref{HJB2}) to derive $K_2^*$. Subsequently, with these elementary estimates, we can compute the mean field state $\bar{x}^*(t)$ from the ODE (\ref{xbar}) given the initial condition. 
 
\vspace{-0.1in}
\subsection{Overall design mechanism}
We are in a position to present a data-driven algorithm to solve Problem \ref{prob1} as described below.
\SetKwComment{Comment}{/* }{ */}
\begin{algorithm}
\caption{data-driven $\varepsilon$-Nash equilibrium computation algorithm  \uppercase\expandafter{\romannumeral1}}\label{alg1}
{\bf Input}: {$K_{1}^0$ such that $A-\rho/2I_n- BK_1^{0}$ is Hurwitz; convergence criterion $\xi$\;
$u_i(t)\gets -K_1^0x_i(t)+\ell_i$(t), $i=1,2,\cdots,N$, $t\in[t_1,t_l]$\;
$P_{11}^0\gets{\bf0}$, $k \gets 0$\;}
\KwData{\!Collect data and compute matrices (\ref{matr1})  and  (\ref{matr2}).}
\KwResult{$u_i^o =-K_1^*x_i(t)-K_2^*\bar{x}^*(t)$, $1\leq i\leq N$.}
\While{$\|P_{11}^{k}-P_{11}^{k-1}\|>\xi$ or $k=0$}{
$k\gets k+1$\;
Update $(P_{11}^k,K_1^k)$ by (\ref{pk})\;
}
$K_1^*\gets K_1^{k}$\;
$P_{11}^* \gets P_{11}^{k}$\;
$B\gets\left(P_{11}^{k}\right)^{-1}(K_1^k)^{\mathrm{T}}R$\;
Solve $A$ and $A+G$ from (\ref{Ai}) and (\ref{Gi}) respectively\;
Solve $P_{12}^*$ from (\ref{HJB2})\;
$K_2^*\gets K_1^*(P_{11}^*)^{-1}P_{12}^*$\;
Solve $\bar{x}^*(t)$ from equation (\ref{xbar})\;
\end{algorithm}
\begin{remark}
Equations (\ref{pk}), (\ref{Ai}), and (\ref{Gi}) are  least-squares problems that aim to minimize the error between the target function and the parameterized function by determining the parameters. These equations can be solved in real-time, provided there is sufficient data and excitation conditions specified in Assumptions \ref{A1}-\ref{A2} are met.  

In the practical realization of Assumptions \ref{A1} and \ref{A2}, exploration noises are injected into the input channels of all agents. Consequently,  the control policies for these agents can be designed as 
\begin{equation}
u_i(t)= -K_1^0x_i(t)+\ell_i(t),~~i=1,2,\cdots,N,\\
\end{equation}
where $K_1^0$ is the feedback gain matrix as stated in Theorem \ref{thm1}, and $\ell_i(t)$, $i=1,2$, denote exploration noises, which can be chosen as stated in Remark \ref{PE_condition}.
\end{remark}
\begin{remark}
In the implementation of the algorithm, it is sufficient to define the error variables using only two specific agents ({\sl e.g., agents $\mathcal{A}_1$ and $\mathcal{A}_2$}). Specifically, one can set $\tilde{x} \!=\! \mathbb{E}[x_1 \!-\! x_2]$ and $\tilde{u} \!=\! \mathbb{E}[u_1 \!-\! u_2]$. Under this configuration, the probing noise applied to each agent can be chosen independently. So there is no need to collect or share state and input information from the rest of the population. Conversely, if all agents share the same probing noise ({\sl i.e., $\ell_1\! =\! ... \!=\! \ell_N$}), then $\mathbb{E}[x_i] = \mathbb{E}[x_{(N)}]$, leading to $\bar{x} = \mathbb{E}[x_1]$ and $\bar{u} = \mathbb{E}[u_1]$. In either case, the data-driven algorithm can be fully executed by gathering trajectories from at most two agents.
\end{remark}
\begin{remark}
We employ the Monte Carlo method with multiple samples to calculate expectations $\tilde{x}(t),\tilde{u}(t)$, $\bar{x}(t)$, and $\bar{u}(t)$. Then, along these trajectories, we can compute data-based matrices (\ref{matr1}), (\ref{matr2}), and (\ref{matr3}). Furthermore, by using the obtained matrices and given a specified convergence criterion, the iterative computation of equation (\ref{pk}) exhibits quadratic convergence within a finite number of steps. It is noted that compared to the data collection phase, the time required to solve (\ref{pk})-(\ref{Gi}) is negligible. 
\end{remark}
\begin{remark}
During the data collection phase, state and input samples are collected from the population until the validity of Assumptions \ref{A1} and \ref{A2} is ensured. Once the algorithm is complete, the feedback gain matrix is updated with the obtained $K_1^*$ and the feedforward component $(K_2^*-K_1^*)\bar{x}(t)$ are merged into the input channels of all agents. Furthermore, in the case of $G={\bf0}$ and $\Gamma=I_n$, equation (\ref{HJB2}) is simplified to a standard ARE. This equation can be approximately solved using an iteration process similar to (\ref{pk}), thus eliminating the need for the calculation of dynamics parameters.
\end{remark}

\section{IRL Algorithm Design: A Special Case Study}\label{sec4}
In this section, we develop an IRL method for a special case where each agent's dynamics and cost function coefficients satisfy specific conditions. Specifically, we consider scenarios where the matrices satisfies
\begin{equation}\label{sym10}
G=\alpha I_n,~~\Gamma = \beta I_n.
\end{equation}
This leads to the conditions $P_{12}G = GP_{12}$ and $Q\Gamma = \Gamma Q$.
Under these conditions, all solutions $P_{12}$ of (\ref{HJB2}) are symmetric matrices. Then, the nonsymmetric ARE is degenerated into a symmetric indefinite ARE. 
\begin{lemma}\label{lem4.1}
Suppose there exist $K_2^k\in\mathbb{R}^{m\times n}$ such that $A-\rho /2I_n-BK_2^0$ and $A+G-\rho/2 I_n-BK_2^0$ are Hurwitz. Let $P_{12}^{k}\in\mathbb{S}^n$ be a solution to the equation
\begin{equation}\label{p12k}
\begin{aligned}
\rho P_{12}^k = &P_{12}^k(A+G-BK_2^{k-1})+(A-BK_2^{k-1})^{\mathrm{T}}P_{12}^k\\
&+\left(K_2^{k-1}\right)^{\mathrm{T}}RK_2^{k-1}+Q-Q\Gamma,
\end{aligned}
\end{equation}
with $K_2^k$ recursively improved by
\begin{equation}\label{k2k}
K_2^k = R^{-1}B^{\mathrm{T}}P_{12}^k,~~k=1,2,\cdots.
\end{equation}
Then, it follows that $\lim \limits_{k\rightarrow\infty}P_{12}^k = P_{12}^*$ and $\lim \limits_{k\rightarrow\infty}K_2^k=K_2^*$.
\end{lemma}

See the proof in Appendix \ref{lem4.1pf}.

To further develop a model-free algorithm, we compute the following differences and integrals:
\begin{equation*}
I_{\tilde{x}\bar{x}}^t \triangleq \!\int_{t}^{t+T}\!\!\!\!\!\!e^{-\rho\tau}\tilde{x}(\tau)\!\otimes\!\bar{x}(\tau)\mathrm{d}\tau,I_{\bar{x}\tilde{x}}^t \triangleq \!\int_{t}^{t+T}\!\!\!\!\!\!e^{-\rho\tau}\bar{x}(\tau)\!\otimes\!\tilde{x}(\tau)\mathrm{d}\tau,
\end{equation*}
\begin{equation*}
I_{\tilde{x}\bar{u}}^t \triangleq \!\int_{t}^{t+T}\!\!\!\!\!\!e^{-\rho\tau}\tilde{x}(\tau)\!\otimes\!\bar{u}(\tau)\mathrm{d}\tau,I_{\bar{x}\tilde{u}}^t \triangleq \!\int_{t}^{t+T}\!\!\!\!\!\!e^{-\rho\tau}\bar{x}(\tau)\!\otimes\!\tilde{u}(\tau)\mathrm{d}\tau,
\end{equation*}
\begin{equation*}
\begin{aligned}
&\delta_{xx}^t \!\triangleq \!\frac{1}{2}e^{-\rho(t+T)}\!(\tilde{x}(t+T)\!\otimes\bar{x}(t+T) \!+\! \bar{x}(t+T)\!\otimes\tilde{x}(t+T))\\
&~~~~~~~-\frac{1}{2}e^{-\rho t}(\tilde{x}(t)\otimes\bar{x}(t)+\bar{x}(t)\otimes\tilde{x}(t)),\\
&~~~~\triangleq [\delta_{1}^t,\delta^t_2,\cdots,\delta_{n^2}^t]^{\mathrm{T}}\in\mathbb{R}^{n^2},
\end{aligned}
\end{equation*}
\begin{equation*}
\delta_{\hat{x}}^t \triangleq [\underbrace{\delta_{n*(i-1)+i},\cdots,\delta_{n*i}}_{i=1,2,\cdots,n}]^{\mathrm{T}}\in\mathbb{R}^{\frac{1}{2}n(n+1)}.
\end{equation*}
We then obtain data-based matrices
\begin{equation}\label{matr3}
\left\{\begin{aligned}
&\Delta_{\hat{x}} = \left[\delta_{\hat{x}}^{t_1},\cdots,\delta_{\hat{x}}^{t_l}\right]^{\mathrm{T}}\in\mathbb{R}^{l\times\frac{1}{2}n(n+1)},\\
&\mathcal{I}_{\tilde{x}\bar{x}}=\left[{I}_{\tilde{x}\bar{x}}^{t_1},\cdots,{I}_{\tilde{x}\bar{x}}^{t_l}\right]^{\mathrm{T}}\in\mathbb{R}^{l\times n^2},\\
&\mathcal{I}_{\bar{x}\tilde{x}}=\left[{I}_{\bar{x}\tilde{x}}^{t_1},\cdots,{I}_{\bar{x}\tilde{x}}^{t_l}\right]^{\mathrm{T}}\in\mathbb{R}^{l\times n^2},\\
&\mathcal{I}_{\tilde{x}\bar{u}}=\left[{I}_{\tilde{x}\bar{u}}^{t_1},\cdots,{I}_{\tilde{x}\bar{u}}^{t_l}\right]^{\mathrm{T}}\in\mathbb{R}^{l\times nm},\\
&\mathcal{I}_{\bar{x}\tilde{u}}=\left[{I}_{\bar{x}\tilde{u}}^{t_1},\cdots,{I}_{\bar{x}\tilde{u}}^{t_l}\right]^{\mathrm{T}}\in\mathbb{R}^{l\times nm}. 
\end{aligned}\right.
\end{equation}
\begin{assume}\label{A4-1}
There exists an $\ell_3>0$ such that for all $l\geq\ell_3$, we have
\begin{equation}\label{rank3}
\mathrm{rank}\left(\left[\mathcal{I}_{\tilde{x}\bar{x}}+\mathcal{I}_{\bar{x}\tilde{x}},\mathcal{I}_{\tilde{x}\bar{u}}+\mathcal{I}_{\bar{x}\tilde{u}}\right]\right)=\frac{1}{2}n(n+1)+mn.
\end{equation}
\end{assume}
The sequence $\{P_{11}^k,K_1^k\}_{k=1}^{\infty}$ generated by iteratively solving equation (\ref{pk}) and its convergence result has been given in Theorem \ref{thm1}.  

\begin{theorem}\label{thm4.1}
Suppose there exists $K_2^0\in\mathbb{R}^{m\times n}$ such that $A-\rho/2 I_n-BK_2^0$ and $A+G-\rho/2 I_n-BK_2^0$ are Hurwitz, and Assumption \ref{A4-1} holds. Then, the sequence $\{P_{12}^k,K_2^k\}_{k=1}^{\infty}$ generated by iteratively solving  
\begin{equation}\label{p12k2}
 \left[\begin{array}{c}
{\mathrm{colm}({P}_{12}^k)}\\
\mathrm{col}(K_2^k)
\end{array}\right]=\left( \left( \Psi_4^k \right)^{\mathrm{T}}\Psi_4^k \right)^{ -1} ( \Psi_4^k )^{\mathrm{T}}\Xi_4^k, 
\end{equation}
where
\begin{equation*}
\left\{\begin{aligned}
&\Psi_4^k = [\Delta_{\hat{x}},-(\mathcal{I}_{\tilde{x}\bar{x}}+\mathcal{I}_{\bar{x}\tilde{x}})(I_n \otimes (K_2^{k-1})^{ \mathrm{T}}R) \\
&~~~~~~~~~-(\mathcal{I}_{\tilde{x}\bar{u}}+\mathcal{I}_{\bar{x}\tilde{u}})(I_n \otimes  R)],\\
&\Xi_4^k=-\mathcal{I}_{\tilde{x}\bar{x}}\mathrm{col}\left((K_2^{k-1})^{\mathrm{T}}RK_2^{k-1}+Q-Q\Gamma\right),
\end{aligned}\right.
\end{equation*}
satisfies $\lim_{k\rightarrow\infty}P_{12}^k=P_{12}^*$ and $\lim_{k\rightarrow\infty}K_2^k=K_2^*$.
\end{theorem}

See the proof in Appendix \ref{thm4.1pf}.

\begin{remark}
{Similar to the analysis in \cite{pang2021robust} and Theorem \ref{thm1}, the quadratic convergence rate of the above iterative process is straightforward to verify. It should be noted that the problem addressed in \cite{xu2023mean} is a special case of the current setting. Specifically, in \cite{xu2023mean}, the scalar coefficients of (\ref{sym10}) are fixed at $\alpha=0$ and $\beta = 1$, which removes the coupling term $Gx_{(N)}$ from the dynamics and reduces the cost functional (\ref{J}) to $J_i\big( u_i,u_{-i} \big) \!= \!\mathbb{E} \Big[ \int_0^{\infty}\!\!\! e^{-\rho\tau} \big(\|x_i- x_{(N)}\|_{Q}^{2} + \|u_i\|_{R}^2 \big) \mathrm{d}\tau \Big]$. Consequently, the development of the model-free method in \cite{xu2023mean} only requires the trajectories of a single agent.}

In contrast, our method necessitates both error trajectories and average trajectories. Furthermore, compared to \cite{xu2023mean}, the current iterative equation involves fewer unknown parameters, thereby reducing the overall computational complexity.
\end{remark}

Finally, substituting $u_i=-K_2^*x_i$ into (\ref{sys1}), we obtain
\begin{equation}\label{x*}
\begin{aligned}
\mathrm{d}x_i(t) =&\left((A-BK_2^*)x_i(t)+Gx_{(N)}(t) \right)\mathrm{d}t\\
&+D\mathrm{d}w_i(t) ,~~i=1,2\cdots,N.
\end{aligned}
\end{equation}
Averaging them yields $\mathrm{d}x_{(N)} = \left(A + G -B K_2^* \right)x_{(N)}+\frac{D}{N}\sum_{i=1}^{N} \mathrm{d}w_i$.
Using the law of large numbers, one has $\lim_{N\rightarrow\infty}x_{(N)}(t) = \bar{x}^*(t)$.
Therefore, the mean field state $\bar{x}^*$ can be approximated by the average state of a large population of agents as described in equation (\ref{x*}).

\section{Numerical Examples}\label{sec5}
In this section, we present two numerical examples to illustrate the effectiveness of our proposed algorithms. The first simulation validates the performance of Algorithm \ref{alg1}. In the second simulation, we examine a special case and compare the results of Algorithms \ref{alg1} and \ref{alg2}.
\subsection{Numerical Example  \uppercase\expandafter{\romannumeral1}}
Consider a large-scale population involving $100$ agents, where each agent's dynamics are described by the matrices
\begin{equation*}
\begin{aligned}
&A=\left[\begin{array}{cc}
5&3\\
10&12
\end{array}\right],~~B=\left[\begin{array}{c}
0\\
1
\end{array}\right],\\
&G=\left[\begin{array}{cc}
1&2\\
3&5
\end{array}\right],~~D=\left[\begin{array}{cc}
0.1&0.1\\
0.1&0.1
\end{array}\right],
\end{aligned}
\end{equation*}
with $x_i\in\mathbb{R}^2$, $u\in\mathbb{R}$, and $w_i$ being a standard two-dimensional Brownian motion. The initial state $x_i(0)$ is uniformly distributed on $[0,2]\times[-1,3]\subset\mathbb{R}^2$ with $Ex_i(0)=[1,1]^{\mathrm{T}}$. 

The parameters of the cost function (\ref{J}) are 
\begin{equation*}
\begin{aligned}
Q=\left[\begin{array}{cc}
1&0\\
0&1
\end{array}\right],~\Gamma=\left[\begin{array}{cc}
1&1\\
1&2
\end{array}\right],~R=0.01,~\rho =0.01.
\end{aligned}
\end{equation*}
Clearly, conditions (A1) and (A2) are satisfied. For comparison, the solutions $P_{11}^*$ and $P_{12}^*$ to AREs (\ref{HJB1})-(\ref{HJB2}) are given by
\begin{equation*}
\begin{aligned}
&P_{11}^*=\!\left[\!\begin{array}{cc}
\!4.1407 & 0.7585\\
\!0.7585 & 0.3843
\end{array}\!\right]\!,\quad \!P_{12}^*=\!\left[\!\begin{array}{cc}
\!3.5326& 1.0085\\
\!0.5717& 0.4275
\end{array}\!\right]\!,
\end{aligned}
\end{equation*}
which result in the following gain matrices
\begin{equation*}
\begin{aligned}
\!\!\!K_1^*\!=\!\left[\!\begin{array}{cc}
\!75.8526 & 38.4335
\end{array}\!\right],\quad K_2^*\!=\!\left[\!\begin{array}{cc}
\!57.1654 & 42.7522
\end{array}\!\right].
\end{aligned}
\end{equation*} 

In this simulation, to implement Algorithm \ref{alg1} using $\mathcal{A}_1$ and $\mathcal{A}_2$, the control inputs are designed as follows
\begin{equation}\label{e1_u1}
\begin{aligned}
{{K_1^0 = [35,25],\quad \ell_i = a_i\sum \nolimits_{j=1}^{100} \sin(w_i^jt),}}
\end{aligned}
\end{equation}
where the amplitudes $a_i$ and the frequencies $w_i^j$, $i,j=1,\cdots,100$, are randomly selected from the ranges $[1,10]$ and $[-100,100]$, respectively. These values are utilize in computing the error variables $\tilde{x}(t)$ and $\tilde{u}$. Furthermore, for calculating the average variables $\bar{x}$ and $\bar{u}$, we set $a_1 = \cdots = a_{100}$ and $w_{1}^j = \cdots = w_{100}^j$, also randomly chosen from $[1, 10]$ and $[-100, 100]$, respectively.
The remaining parameters are configured as follows: $T=T_s=0.1\text{~[sec]}$, $[t_1,t_l]=[0,5]\text{~[sec]}$, and the convergence criterion $\xi=10^{-6}$. 

\begin{figure*}[!htb]
\centering
\includegraphics[scale=0.26]{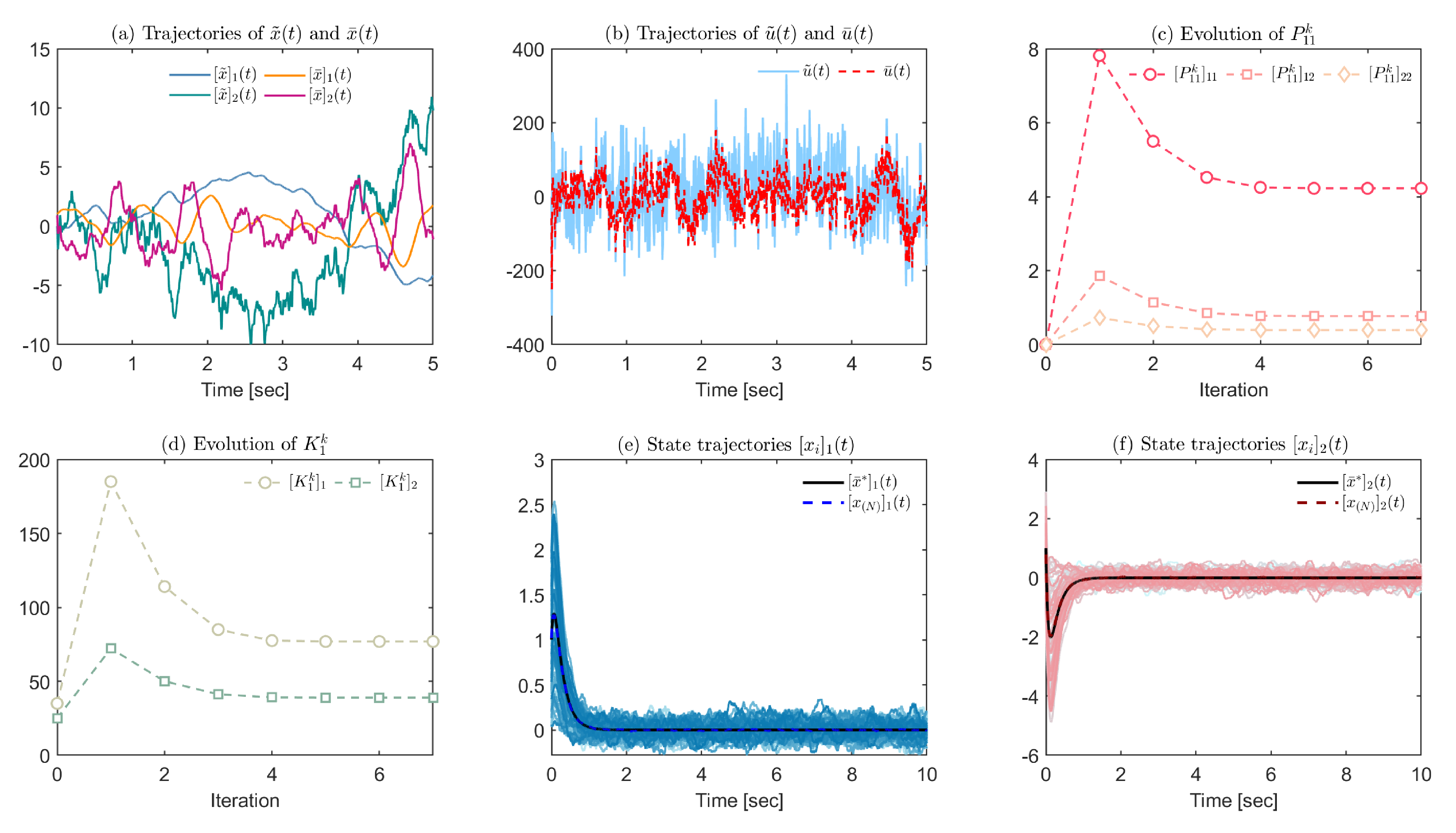}
\centering
\caption{(a) and (b): Data used for computing matrices (\ref{matr1}), (\ref{matr2}), and (\ref{matr3}); (c) and (d): Simulation results of Algorithm \ref{alg1}; (e) and (f): State trajectories of the population with the obtained decentralized control policies (\ref{uo}).}\label{e1r1}
\end{figure*}

We collect 100 state and input samples from the selected two agents to approximate the error variables $\tilde{x}(t)$ and $\tilde{u}(t)$, as well as the average variables $\bar{x}(t)$ and $\bar{u}(t)$ (see Fig. \ref{e1r1} (a)-(b) for these trajectories used in Algorithm \ref{alg1}), then compute the data-based matrices until rank conditions (\ref{rank1}) and (\ref{rank2}) are satisfied. The simulation was conducted using data obtained from the system at every $T_s=0.1$ [sec], with the integrals approximated using the rectangle rule for numerical integration. 

\SetKwComment{Comment}{/* }{ */}
\begin{algorithm}
\caption{Data-driven $\varepsilon$-Nash equilibrium computation algorithm  \uppercase\expandafter{\romannumeral2} }\label{alg2}
{\bf Input}: {$K_{1}^0$ such that $A-\rho/2I_n-BK_1^{0}$ is Hurwitz, convergence criterion $\xi$\;
$u_i(t)\gets -K_1^0x_i(t)+\ell_i$(t), $i=1,2$, $t\in[t_1,t_l]$\;
$P_{11}^0\gets{\bf0}$, $P_{12}^0\gets{\bf0}$, $k \gets 0$\;}
\KwData{\!Collect data and compute matrices (\ref{matr1}) and (\ref{matr3}).}
\KwResult{$u_i^o =-K_1^*x_i(t)-K_2^*\bar{x}^*(t)$, $1\leq i\leq N$.}
\While{$\|P_{11}^{k}-P_{11}^{k-1}\|>\xi$ or $k=0$}{
$k\gets k+1$\;
Update $(P_{11}^k,K_1^k)$ by (\ref{pk})\;
}
$K_1^*\gets K_1^{k}$\;
$k \gets 0$\;
\While{$\|P_{12}^{k}-P_{12}^{k-1}\|>\xi$ or $k=0$}{
$k\gets k+1$\;
Update $(P_{12}^k,K_2^k)$ by (\ref{p12k2})\;
}
$K_2^*\gets K_2^{k}$\;
$u_i\gets -K_2x_i$\;
Collect state trajectories $x_i(t)$ and calculate the average state $x_{(N)}(t)$\;
$\bar{x}(t)^*\gets x_{(N)}(t)$\;
\end{algorithm}
\begin{table}[!htb]
 \caption{Approximate error values.}\label{e1tab1}
\begin{tabular}{|c|c|c|c|}
\hline
Error & Value & Error & Value\\ \hline
$\|P_{11}^*-{P}_{11}^7\|_2$ & $0.0868$ &$\|A-\hat{A}\|_2$ & $0.0673$\\
$\|K_{1}^*-{K}_{1}^7\|_2$ & $1.1585$& $\|B-\hat{B}\|_2$ & $7.2635\times 10^{-4}$\\
$\|P_{12}^*- \hat{P}_{12}\|_2$ & $0.0263
$  &$\|A_G-\hat{A}_{G}\|_2$ & $0.0982$\\
$\|K_{2}^*-\hat{K}_{2}\|_2$ & $0.3430$ &$\|G-\hat{G}\|_2$ & $0.0594$\\
\hline
\end{tabular}\vspace{-2ex}
\end{table}
The convergence result of the sequence ${P_{11}^k,K_1^k}$ is shown in Fig. \ref{e1r1} (c)-(d). From the simulation, the sequence $\{P_{11}^k,K_1^k\}$ converges at the $7$-th iteration under the convergence criterion $\xi$, with final values  
\begin{equation}
\begin{aligned}
\!\!\!{P}_{11}^7 = \left[\!\begin{array}{cc}
\!4.2254 & 0.7715\\
\!0.7715 & 0.3898
\end{array}\!\right] ,~~{K}_1^7 = \left[ \begin{array}{c}
\!76.8957\\
\!38.9376
\end{array}\right]^{\mathrm{T}}.
\end{aligned}
\end{equation}
Furthermore, using the above results and data-based matrices (\ref{matr1})-(\ref{matr2}), the drift coefficients are identified as 
\begin{equation*}
\begin{aligned}
&\hat{B}=\left[\begin{array}{c}
-0.0007\\
1.0003
\end{array}\right],\quad  \hat{A}=\left[\begin{array}{cc}
\!4.9889 & 2.9894\\
\!9.9945 & 12.0663
\end{array}\right],\\
&\hat{A}_{G}=\left[\begin{array}{cc}
\!5.9904 & 4.9797\\
\!12.9357 & 17.0736
\end{array}\right],\quad \hat{G}=\left[\begin{array}{cc}
\!1.0015 & 1.9902\\
\!2.9411 & 5.0073
\end{array}\right],
\end{aligned}
\end{equation*}
where $A_G\triangleq A+G$, the notation $\hat{\cdot}$ is used to denote the estimated value of matrix coefficients, enabling differentiation from the true values.

Substituting the obtained approximations into equations (\ref{HJB2}) and (\ref{xbar}) yields
\begin{equation}
\!\!\!\hat{P}_{12}=\left[\begin{array}{cc}
\!3.5582&1.0131\!\\
\!0.5746&0.4293\!
\end{array}\right],\quad \hat{K}_2=\left[\begin{array}{c}
\!57.4598\!\\
\!42.9284\!
\!\end{array}\right]^{\mathrm{T}},
\end{equation}
and substituting into (\ref{xbar}) yields the mean field state trajectory $\bar{x}^*(t)$. All approximation errors are summarized in Tab. \ref{e1tab1}.

Finally, the decentralized control policies are applied to the population, and state trajectories of the population are shown in Fig. \ref{e1r1} (e)-(f). It can be seen that the average state $x_{(N)}$ of the population is close to the mean field state trajectory $\bar{x}^*(t)$, verifying the consistency condition.

\subsection{Numerical Example  \uppercase\expandafter{\romannumeral2}}
In this example, we focus on a special case (as described in Section \ref{sec4}) to verify the effectiveness of Algorithm \ref{alg2}. The dynamics of each agent are given by the following matrices
\begin{equation*}
\begin{aligned}
A\!=\!\left[\begin{array}{cccc}
-5&1&0&1\\
2&-4&1.5&0\\
0&0&-2.5&1\\
1&0&1&-5
\end{array}\right],B\!=\!\left[\begin{array}{c}
0\\
0\\
0\\
1
\end{array}\right],D\!=\!\left[\begin{array}{c}
0\\
0\\
0\\
0.01
\end{array}\right],
\end{aligned}
\end{equation*}
and $G= \text{diag}(-0.9,-0.9,-0.9,-0.9)$.
Here, $x_i\in\mathbb{R}^4$, $u\in\mathbb{R}$, and $w_i$ is a standard one-dimensional Brownian motion. The initial state $x_i(0)$ is uniformly distributed on $[0,10]\times [0,10]\times[0,10]\times [0,10]\subset\mathbb{R}^4$ with $Ex_i(0)=[5,5,5,5]^{\mathrm{T}}$. The number of agents is $100$. Additionally, the parameters of the cost function (\ref{J}) are $R=\rho=0.1$, 
$Q = \text{diag}(50,50,50,50)$ and $\Gamma= \text{diag}(0.9,0.9,0.9,0.9)$.
\begin{figure*}[!htb]
\centering
\includegraphics[scale=0.21]{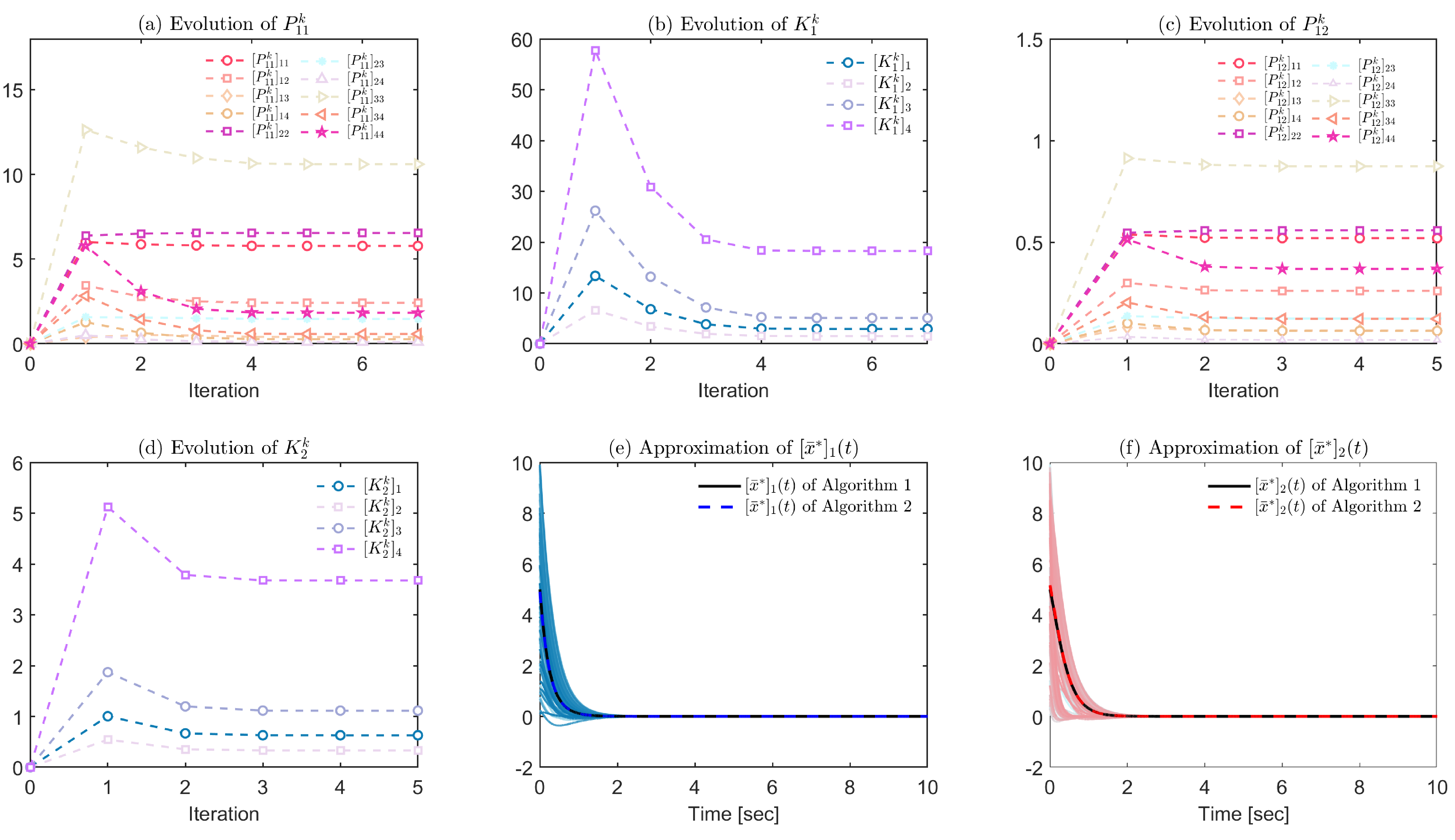}
\centering
\caption{(a)(b): data used for computing matrices (\ref{matr1}) and (\ref{matr2}); (c)(d): simulation results of Algorithm \ref{alg1}; (e)(f): state trajectories of the population with the obtained decentralized strategies.}\label{e2r1}
\end{figure*}

We have verified that conditions (A1) and (A2) are satisfied. Moreover, it is obvious that the matrices $A$ and $A+G$ are Hurwitz. To implement Algorithms \ref{alg1} and \ref{alg2}, the feedback gain matrix for the designed inputs are chosen as $K_1^0=[0,0,0,0]$. The exploration noises are selected as
\begin{equation*}
\begin{aligned}
\ell_i = a_i\sum \nolimits_{j=1}^{100}\sin(w_i^jt), ~~i=1,2,\cdots,100,
\end{aligned}
\end{equation*}
where $a_i$ and $w_i^j$ are randomly selected from $[1,10]$ and $[-100,100]$, respectively. These parameters are used to compute $\tilde{x}$ and $\tilde{u}$. Next, the coefficients are reset to $a_1=\cdots=a_{100}$ and $w_1^j=\cdots=w_{100}^j$, which are still selected randomly from the same ranges, to calculate $\bar{x}$ and $\bar{u}$.
The other parameters are set as follows: $T_s=10^{-3}\text{~[sec]}$, $T=0.1\text{~[sec]}$, $[t_1,t_l]=[0,20]\text{~[sec]}$, and $\xi=10^{-6}$. 

We collect $100$ sate and input samples of $\mathcal{A}_1$ and $\mathcal{A}_2$, respectively, to compute the data-based matrices until rank conditions (\ref{rank1}), (\ref{rank2}), and (\ref{rank3}) are satisfied,  and then implement Algorithm \ref{alg1} and Algorithm \ref{alg2}.

The convergence results of the sequences $\{P_{11}^k,K_1^k\}$ and $\{P_{12}^k,K_2^k\}$ are shown in Fig. \ref{e2r1} (a)-(d). From the simulation, the sequence $\{P_{11}^k,K_1^k\}$ converges at the $7$-th iteration, and the sequence $\{P_{12}^k,K_2^k\}$ is convergent at the $5$-th under $\xi=10^{-6}$. The final values are
\begin{equation*}
\begin{aligned}
\!\!\!\!{P}_{11}^7 \!\!=\! \!\left[\!\begin{array}{cccc}
\!5.7761&\!2.4144&\!0.3968&\!0.2694\\
\!2.4144&\!6.5390&\!1.4569&\!0.1018\\
\!0.3968&\!1.4569&\!10.5962&\!0.5727\\
\!0.2694&\!0.1018&\!0.5727&\!1.8280\\
\end{array}\!\right]\!, \!{K}_1^7\!\! =\!\! \left[ \begin{array}{c}
\!2.9030\!\\
\!1.4859\!\\
\!5.0644\!\\
\!18.2495\!
\end{array}\right]^{\!\mathrm{T}}\!\!\!\!,
\end{aligned}
\end{equation*}
\begin{equation*}
\begin{aligned}
\!\!\!\!{P}_{12}^5 \!\!= \!\!\left[\!\begin{array}{cccc}
\!0.5193&\!0.2598&\!0.0643&\!0.0632\\
\!0.2598& \!0.5577&\!0.1242&\!0.0186\\
\!0.0643&\!0.1242&\!0.8735&\!0.1223\\
\!0.0632&\!0.0186&\!0.5727&\!0.3683\\
\end{array}\!\right]\!, \!{K}_2^5 \!\!=\!\! \left[ \begin{array}{c}
\!0.6285\!\\
\!0.3299\!\\
\!1.1132\!\\
\!3.6745\!
\end{array}\right]^{\!\mathrm{T}}\!\!\!\!.
\end{aligned}
\end{equation*}

Next, we set $K_2^*=K_2^5$, change the control input to $u_i=-K_2^*x_i$, and collect samples of the population to approximate $\bar{x}^*(t)$. The state trajectory of each agent and the approximated mean field state trajectory are shown in Fig. \ref{e2r1} (e)-(f).

Furthermore, we implement Algorithm \ref{alg1} to obtain the drift coefficients  
\begin{equation*}
\begin{aligned}
\hat{B}=\left[\begin{array}{cccc}
5.3051\times10^{-4}&0.0086&-0.0075&1.0001
\end{array}\right]^{\mathrm{T}},
\end{aligned}
\end{equation*}
\begin{equation*}
\begin{aligned}
\hat{A}=\left[\begin{array}{cccc}
\!-4.9366 & 1.0055 & -0.0394 & 0.9983\\
\!2.0090  & -4.0090 & 1.5049 & -0.0018\\
\!0.0759  & 8.035\times10^{-4} & -2.5411 & 0.9970\\
\!1.3076  & 0.0896 & 0.7418 & -4.9984
\end{array}\right], 
\end{aligned}
\end{equation*}
\begin{equation*}
\begin{aligned}
\hat{G}=\left[\begin{array}{cccc}
\!-0.9601&-0.0194&0.0488&7.3103\times10^{-4}\\
\!0.0018&-0.9040&0.0013&2.381\times10^{-6}\\
\!-0.0538&-0.0192&-0.8558&8.2736\times10^{-4}\\
\!-0.2028&	-0.0442& 0.1419&-0.8947
\end{array}\right],
\end{aligned}
\end{equation*}
\begin{equation*}
\begin{aligned}
\hat{A}_{G}=\left[\begin{array}{cccc}
\!-5.8968&0.9861&0.0093&0.9990\\
\!2.0108&-4.9130&1.5062&-0.0017\\
\!0.0221&-0.0184&-3.3969&0.9979\\
\!1.1048&0.0454&0.8837&-5.8931 
\end{array}\right].
\end{aligned}
\end{equation*}
Then, solving (\ref{HJB2}) yields
\begin{equation*}
\begin{aligned}
\!\!\!\!{P}_{12}^5 \!\!= \!\!\left[\!\begin{array}{cccc}
\!0.5333&\!0.1745&\!0.0866&\!0.0796\\
\!0.1742& \!0.5917&\!0.1307&\!0.0293\\
\!0.0710&\!0.1231&\!0.9024&\!0.1070\\
\!0.0723&\!0.0268&\!0.1128&\!0.3667\\
\end{array}\!\right]\!, \!{K}_2^5 \!\!=\!\! \left[ \begin{array}{c}
\!0.7230\!\\
\!0.2678\!\\
\!1.1280\!\\
\!3.6666\!
\end{array}\right]^{\!\mathrm{T}}\!\!\!\!, 
\end{aligned}
\end{equation*}
Finally, solving the ODE \ref{xbar} gives $\bar{x}^*(t)$, as shown in \ref{e2r1} (e)-(f). The approximations of the mean field state trajectory by both algorithms are nearly identical.  

Approximation errors of both algorithms are concluded in Tab. \ref{e1tab2}, confirming that all approximations are close to the corresponding true values.
\begin{table}[!htb]
 \caption{Approximate error values.}\label{e1tab2}
\begin{tabular}{|cc|cc|}
\hline
\multicolumn{2}{|c|}{Algorithm \ref{alg1}}     & \multicolumn{2}{c|}{Algorithm \ref{alg2}}     \\ \hline
\multicolumn{1}{|c|}{Error} & Value & \multicolumn{1}{c|}{Error} & Value \\ \hline
\multicolumn{1}{|c|}{$\|P_{11}^*-P_{11}^{7}\|_2$} & $0.5814$ & \multicolumn{1}{c|}{$\|P_{11}^*-P_{11}^{7}\|_2$} &  $0.5814$ \\ \hline
\multicolumn{1}{|c|}{$\|K_1^*-K_1^{7}\|_2$} & $0.1377$ & \multicolumn{1}{c|}{$\|K_1^*-K_1^{7}\|_2$} &  $0.1377$ \\ \hline
\multicolumn{1}{|c|}{$\|P_{12}^*-P_{12}^{5}\|_2$} & $0.1099$ & \multicolumn{1}{c|}{$\|P_{12}^*-\hat{P}_{12}\|_2$} &  $ 0.0238$ \\ \hline
\multicolumn{1}{|c|}{$\|K_{2}^*-K_{2}^{5}\|_2$} & $0.0955$ & \multicolumn{1}{c|}{$\|K_{2}^*-\hat{K}_{2}\|_2$} & $0.0339$  \\ \hline
\end{tabular}
\end{table}

\section{Conclusion}\label{sec6}
In general, a key point of model-free control design is how to avoid using agents' dynamic model information. According to the adaptive control policy, the methodology can be categorized as explicit and implicit adaptive approaches. The former first identifies the model and then calculates the control gain with the identified model, while the latter directly adjusts the control gain without model information. In this sense, the first design method proposed in this paper for general LQG games can be considered as a mixed approach. This approximate Nash equilibrium computation method proposed in this paper is based on AREs (\ref{HJB1}) and (\ref{HJB2}). Benefiting from the standard structure of (\ref{HJB1}), the feedback gain of the agent’s own state is obtained through a policy iteration approach, which is achieved by technically introducing the error system of two randomly selected agents. Subsequently, based on the obtained solutions, the model parameters of the agent dynamics are identified by again using the collected sampling data of the selected agents. As with explicit adaptive design, the mean field feedforward gain is computed by solving the nonsymmetric ARE (\ref{HJB2}). Furthermore, in special cases, the second method employs IRL techniques to approximate the solutions of (\ref{HJB1}) and (\ref{HJB2}), similar to the implicit adaptive design methodology. In future work, we plan to expand the scope of the IRL technique beyond the LQG framework to cover more general cases, wherein the approximate Nash equilibria arise from the solution of forward-backward partial differential equations.

\bibliographystyle{IEEEtran}        

\appendix
\subsection{Proof of Lemma \ref{lem4.1}}\label{lem4.1pf}
Starting from (\ref{sym10}), we can rearrange equation (\ref{p12k}) as
\begin{equation}\label{p12k'}
\begin{aligned}
{\bf0} \!=\!  P_{12}^{k}\tilde{A}_{k-1}\!+\!\tilde{A}_{k-1}^{\mathrm{T}}P_{12}^{k}+\!\left(K_2^{k-1}\right)^{\mathrm{T}}\!\!RK_{2}^{k-1}\!+\!Q\!-\!Q\Gamma,
\end{aligned}
\end{equation}
where $\tilde{A}_{k}\triangleq A + 1/2G-\rho/2 I_n-BK_2^{k}$. Equation (\ref{p12k}) is equivalent to (\ref{p12k'}). We will proof the lemma based on (\ref{p12k'}).

We first show that $P_{12}^{k}\geq P_{12}$ and the Hurwitz property of $\tilde{A}_{k-1}$ hold for all $k\in\mathbb{N}_+$ by mathematical induction.

i) For $k=1$, by the initial condition that $A-\rho/2 I_n-BK_2^0$ and $A+G-\rho/2 I_n-BK_2^0$ are Hurwitz, it is easy to get $\tilde{A}_0$ is Hurwitz.

To show the inequality holds, rewrite equation (\ref{HJB2}) with $P_{12}=P_{12}^*$, subtract and add $(K_2^0)^{\mathrm{T}}B^{\mathrm{T}}P_{12}^*$ and $P_{12}^*BK_{2}^0$, and substitute (\ref{k2k}) into it to obtain
\begin{equation}\label{lem2pfeq1}
\begin{aligned}
{\bf0} =& P_{12}^*\tilde{A}_0\!+\!\tilde{A}_0^{\mathrm{T}}P_{12}^*\!+\!\left(K_2^*\right)^{\mathrm{T}}\!\!RK_2^0\!+\!\left(K_2^0\right)^{\mathrm{T}}\!\!RK_{2}^*\\
&-\left(K_2^*\right)^{\mathrm{T}}\!\!RK_2^*+Q-Q\Gamma.
\end{aligned}
\end{equation}
Subtract equation (\ref{lem2pfeq1}) from equation (\ref{p12k'}) with $k=1$, yielding
\begin{equation*} 
\begin{aligned}
{\bf0}=&(P_{12}^1-P_{12}^*)\tilde{A}_0+\tilde{A}_0^{\mathrm{T}}(P_{12}^1-P_{12}^*)\\
&+(K_2^*-K_2^0)^{\mathrm{T}}R(K_2^*-K_2^0).
\end{aligned}
\end{equation*}
From the Hurwitz property of $\tilde{A}_0$ and \cite[Proposition 8.13.1]{lancaster1985theory}, we have 
\begin{equation*}
\begin{aligned}
P_{12}^{1}\!-\!P_{12}^*\!=\!\int_0^{\infty}\!\!\!e^{\tilde{A}_0\tau}(K_2^*-K_2^0)^{\mathrm{T}}R(K_2^*-K_2^0) e^{\tilde{A}_0^{\mathrm{T}}\tau}\mathrm{d}\tau,
\end{aligned}
\end{equation*}
which implies $P_{12}^1\geq P_{12}^*$.
 
ii) Suppose the Hurwitz property of $\tilde{A}_{k-1}$ holds for $k>1$, we prove that it also holds for $k+1$ and that the inequality $P_{12}^{k}\geq P_{12}^*$ holds for $k>1$. 

Rewrite equation (\ref{lem2pfeq1}) as
\begin{equation*} 
\begin{aligned}
{\bf0} = & P_{12}^*\tilde{A}_{k-1}\!+\!\tilde{A}_{k-1}^{\mathrm{T}}P_{12}^*+\left(K_2^*\right)^{\mathrm{T}}\!RK_2^{k-1}\\
&+\!\left(K_2^{k-1}\right)^{\mathrm{T}}\!\!RK_{2}^*-\left(K_2^*\right)^{\mathrm{T}}\!\!RK_2^*+Q-Q\Gamma.
\end{aligned}
\end{equation*}
Subtract this from equation (\ref{p12k'}) to get
\begin{equation}\label{lem2pfeq3}
\begin{aligned}
{\bf0}= &(P_{12}^k-P_{12}^* )\tilde{A}_{k-1}+\tilde{A}_{k-1}^{\mathrm{T}}(P_{12}^k-P_{12}^*)\\
&+(K_s-K_s^{k-1})^{\mathrm{T}}R(K_s-K_s^{k-1}).
\end{aligned}
\end{equation}
Since $\tilde{A}_{k-1}$ is Hurwitz and $R>0$, we have $P_{12}^k\geq P_{12}^*$.
 
To prove $\tilde{A}_{k}$ is Hurwitz, transform equation (\ref{lem2pfeq3}) into 
\begin{equation*} 
\begin{aligned}
\!\!\!{\bf0}=&(P_{12}^k-P_{12}^*)\tilde{A}_{k}\!+\!\tilde{A}_{k}^{\mathrm{T}}(P_{12}^k-P_{12}^*)\!+\!(K_2^k\!-\!K_2^{k-1})^{\!\mathrm{T}}R\\
\!\!\!&\times (K_2^k-K_2^{k-1})+(K_2^k-K_2^*)^{\mathrm{T}}R(K_2^k-K_2^*).
\end{aligned}
\end{equation*}
Assume $\tilde{A}_{k}x=\lambda x$ for some $\lambda$ with $\mathrm{Re}(\lambda)\geq0$ and vector $x\neq{\bf0}$. Then, we have
\begin{equation}\label{lem2pfeq5}
\begin{aligned}
\!\!\!\!\!&0=2\lambda x^{\mathrm{T}}(P_{12}^k-P_{12}^*)x+x^{\mathrm{T}}(K_2^k\!-\!K_2^{k-1})^{\!\mathrm{T}}R(K_2^k\\
\!\!\!\!\!&~~~~~-K_2^{k-1})x\!+\!x^{\mathrm{T}}(K_2^k-K_2^*)^{\mathrm{T}}R(K_2^k-K_2^*)x.
\end{aligned}
\end{equation}
Since $P_{12}^k-P_{12}^*\geq0$, and $\!(K_2^{k}-K_2^{k-1})^{\!\mathrm{T}} R(K_2^{k}\!-\!K_2^{k-1})\!+\!(K_2^{k} - K_2^*)^{\!\mathrm{T}} R(K_2^k-K_2^*)\geq0$, equation (\ref{lem2pfeq5}) implies 
\begin{equation*}
x^{\mathrm{T}}(K_2^k-K_2^{k-1})^{\mathrm{T}}R(K_2^k-K_2^{k-1})x=0,
\end{equation*}
which gives $(K_2^{k}-K_2^{k-1})x={\bf0}$. Consequently, we get $\tilde{A}_kx=\tilde{A}_{k-1}x=\lambda x$. This contradicts the induction assumption. 
Thus, the inequality $P_{12}^k\geq P_{12}^*$ and the Hurwitz property of $\tilde{A}_{k-1}$ hold all $k\in\mathbb{N}_+$. Next, we show the inequality $P_{12}^k\geq P_{12}^{k+1}$ also holds for all $k\in\mathbb{N}_+$.

Next, we show the inequality $P_{12}^k\geq P_{12}^{k+1}$ holds for all $k\in\mathbb{N}_+$. 
Rewrite equation (\ref{p12k'}) as
\begin{equation}\label{sr1}
\begin{aligned}
{\bf0} =& P_{12}^{k}\tilde{A}_{k}+\tilde{A}_{k}^{\mathrm{T}}P_{12}^{k}\!+\!\left(K_2^{k}\right)^{\mathrm{T}}RK_{2}^{k}\!+\!\left(K_2^{k}-K_{2}^{k-1}\right)^{\mathrm{T}}\!\\
&\times R\left(K_2^{k}-K_{2}^{k-1}\right)+\!Q\!-\!Q\Gamma.
\end{aligned}
\end{equation} 
Replace $k$ in (\ref{p12k'}) by $k+1$, yielding
\begin{equation}\label{sr2}
\begin{aligned}
{\bf0} =& P_{12}^{k+1}\tilde{A}_{k}+\tilde{A}_{k}^{\mathrm{T}}P_{12}^{k+1}+\left(K_2^{k}\right)^{\mathrm{T}}RK_{2}^{k}\!+\!Q\!-\!Q\Gamma,
\end{aligned}
\end{equation}
Subtract (\ref{sr2}) from (\ref{sr1}) to obtain
\begin{equation*}
\begin{aligned}
{\bf0} =& (P_{12}^{k}-P_{12}^{k+1})\tilde{A}_{k}+\tilde{A}_{k}^{\mathrm{T}}(P_{12}^{k}-P_{12}^{k+1})\!+\!\left(K_2^{k}\right)^{\mathrm{T}}RK_{2}^{k}\\
&+\!\left(K_2^{k}-K_{2}^{k-1}\right)^{\mathrm{T}}R\left(K_2^{k}-K_{2}^{k-1}\right),
\end{aligned}
\end{equation*}
which implies that $P_{k+1}\geq P_{k}$, $k\in\mathbb{N}_+$.
 
Thus, the sequence $\{P_{12}^k\}_1^{\infty}$ is monotonically decreasing and bounded below by $P_{12}^*$. Consequently, it converges to some limit. Since $P_{12}^*$ satisfies equation (\ref{p12k}) with $K_2^{k-1}=K_2^*$, we have that $\lim_{k\rightarrow\infty}P_{12}^k=P_{12}^*$. Given that $K_2^k$ is the unique solution of (\ref{k2k}), the convergence of $\{K_2^k\}_1^{\infty}$ follows directly from the convergence of $\{P_{12}^k\}_1^{\infty}$. Consequently, we also have $\lim_{k\rightarrow\infty}K_s^k=K_s$.

Hence, the proof is complete.

\subsection{Proof of Theorem \ref{thm4.1}}\label{thm4.1pf}
Define the function
\begin{equation}\label{v4}
V_4(t,\tilde{x},\bar{x})= e^{-\rho t}\tilde{x}^{\mathrm{T}}P_{12}^k\bar{x}.
\end{equation}
The time derivative of $V_4(t)$ is
\begin{equation}\label{dV4}
\begin{aligned}
&\mathrm{d}V_4(t) = e^{-\rho t}\Big(\!\!-\rho \tilde{x}^{\mathrm{T}}(t)P_{12}^k\bar{x}(t)+\big(A\tilde{x}(t)+B\tilde{u}(t)\big)^{\mathrm{T}}\\
&\times P_{12}^k\bar{x}(t)+\tilde{x}^{\mathrm{T}}(t)P_{12}^k\big((A+G)\bar{x}(t)+B\bar{u}(t)\big)\!\!\Big)\mathrm{d}t
\end{aligned}
\end{equation}
Substitute (\ref{p12k}) and (\ref{k2k}) into the above expression and integrate over the interval $[t,t+T)$ to obtain
\begin{equation}\label{iV4}
\begin{aligned}
&e^{-\rho(t+T)}\tilde{x}^{\mathrm{T}}(t+T)P_{12}^k\bar{x}(t+T)-e^{\rho t}\tilde{x}^{\mathrm{T}}(t)P_{12}^k\bar{x}(t)\\
&=\int_t^{t+T}e^{–\rho\tau}\left(K_2^{k-1}\tilde{x}(\tau)+\tilde{u}(\tau)\right)^{\mathrm{T}}RK_2^{k}\bar{x}(\tau)\mathrm{d}\tau\\
&+\int_t^{t+T}e^{–\rho\tau}\left(K_2^{k-1}\bar{x}(\tau)+\bar{u}(\tau)\right)^{\mathrm{T}}RK_2^{k}\tilde{x}(\tau)\mathrm{d}\tau\\
&-\!\!\int_t^{t+T}\!\!\!\!\!\!\!\!\!e^{-\rho\tau}\bar{x}^{\mathrm{T}}(\tau)\!\left(\!\left({K_2^{k-1}}\!\right)^{\!\mathrm{T}}RK_2^{k-1}\!+\!Q-Q\Gamma\!\right)\tilde{x}(\tau)\mathrm{d}\tau.
\end{aligned}
\end{equation}
Given the symmetric property of $P_{12}^k$ and the Kronecker product representation, weh have
\begin{equation*}
\begin{aligned}
&\delta_{\hat{x}}^t{\mathrm{colm}(P_{12}^k)}=\left(I_{\tilde{x}\bar{x}}^{t}+I_{\bar{x}\tilde{x}}^{t}\right)^{\mathrm{T}}\left(I_{n}\otimes \left(RK_{2}^{k-1}\right)^{\mathrm{T}}\right)\mathrm{col}(K_2^k)\\
&\!+\!\left(I_{\!\bar{x}\tilde{u}}^{\!t}\!+\!I_{\!\tilde{x}\bar{u}}^t\right)^{\!\mathrm{T}}\!\!\!\left(I_{n}\!\otimes\!R\right)\mathrm{col}(K_2^k)\!-\!I_{\bar{x}\tilde{x}}^{t}\Big(\left(K_{2}^{k-1}\right)^{\mathrm{T}}RK_{2}^{k-1}\\
&+\!Q\!-\!Q\Gamma\Big).
\end{aligned}
\end{equation*}
By combing $l$ sets of data and assuming the inverse of $(\Psi_4^k)^{\mathrm{T}}\Psi_4^k$ exists, we derive equation (\ref{p12k2}). 

The next step is to show that $\Psi_4^k$ has full column rank for all $K\in\mathbb{N}_+$, which implies the existence of the inverse matrix and thus the uniqueness of the solution of equation (\ref{p12k2}). We prove it by contradiction. 

Assume $\Psi_4^kX=0$ for a nonzero $X=[X_1^{\mathrm{T}},X_2^{\mathrm{T}}]^{\mathrm{T}}$, where $X_1\in\mathbb{R}^{\frac{1}{2}n(n+1)}$ and $X_2\in\mathbb{R}^{nm}$. A symmetric matrix $Y\in\mathbb{S}^n$ is uniquely determined by {$\mathrm{colm}(Y)=X_1$}, and a matrix $Z\in\mathbb{R}^{m\times n}$ is determined by $\mathrm{col}(Z)=X_2$.

Using equations (\ref{dV4}) and (\ref{iV4}), we get
\begin{equation}\label{p4eq1}
\!\Psi_4^k X \!=\! \frac{1}{2}\left(\mathcal{I}_{\tilde{x}\bar{x}}+\mathcal{I}_{\bar{x}\tilde{x}}\right)\mathrm{col}(M)+\left(\mathcal{I}_{\tilde{x}\bar{u}}+\mathcal{I}_{\bar{x}\tilde{u}}\right)\mathrm{col}(N),
\end{equation} 
where $M=-\rho Y+Y(A+\frac{1}{2}G-BK_2^{k-1})+(A+\frac{1}{2}G-BK_2^{k-1})^{\mathrm{T}}Y+(K_{2}^{k-1})^{\mathrm{T}}N+N^{\mathrm{T}}K_2^{k-1}$ and $N=B^{\mathrm{T}}Y-RZ$.
Since $M\in\mathbb{S}^n$, letting $\mathcal{I}_{\hat{\tilde{x}}=[I_{\hat{\tilde{x}}}^{t_1},\cdots,I_{\hat{\tilde{x}}}}^{t_l}]$, one has
\begin{equation}\label{p4eq2}
\frac{1}{2}\left(\mathcal{I}_{\tilde{x}\bar{x}}+\mathcal{I}_{\bar{x}\tilde{x}}\right)\mathrm{col}(M)=\mathcal{I}_{\hat{\tilde{x}}}{\mathrm{colm}(M)}, 
\end{equation} 
where $I_{\hat{x}}^t \triangleq [\underbrace{I_{n*(i-1)+i},\cdots,I_{n*i}}_{i=1,2,\cdots,n}]^{\mathrm{T}}\in\mathbb{R}^{\frac{1}{2}n(n+1)}$, $I_{xx}^t = \frac{1}{2}\int_t^{t+T}e^{-\rho\tau}(\tilde{x}(\tau)\otimes\bar{x}(\tau)+\bar{x}(\tau)\otimes\tilde{x}(\tau))\mathrm{d}\tau\triangleq [I_{1}^t,I^t_2,\cdots,I_{n^2}^t]^{\mathrm{T}}\in\mathbb{R}^{n^2}$. 

Under Assumption \ref{A4-1}, the matrix $[\mathcal{I}_{\hat{\tilde{x}}},\mathcal{I}_{\tilde{x}\bar{u}}+\mathcal{I}_{\bar{x}\tilde{u}}]$ has full column rank. Hence, from $\Psi_k^4X={\bf0}$ and equations (\ref{p4eq1})-(\ref{p4eq2}), we get ${\mathrm{colm}(M)}={\bf 0}$ and $\mathrm{col}(N)={\bf 0}$. 

This result in $-\rho Y+Y(A+\frac{1}{2}G-BK_2^{k-1})+(A+\frac{1}{2}G-BK_2^{k-1})^{\mathrm{T}}Y={\bf0}$. Since $A-\rho/2I_n+1/2G-BK_2^{k-1}$ is Hurwitz, it implies $Y={\bf0}$. Consequently, $X_2={\bf0}$. This leads to $X={\bf0}$, which contradicts the assumption that $X$ is a nonzero vector. Therefore,  $\Psi_4^k$ must have full column rank for all $k\in\mathbb{N}_+$.

Hence, the proof is complete.

\end{document}